\documentclass[12pt,a4paper,notitlepage]{article}

\usepackage{amssymb,amsmath,eucal}
\usepackage[dvips]{lscape,graphicx}
\usepackage{cite}

\newcommand{\bc}{\begin{center}}
\newcommand{\ec}{\end{center}}
\newcommand{\bd}{\begin{displaymath}}
\newcommand{\ed}{\end{displaymath}}
\newcommand{\be}{\begin{equation}}
\newcommand{\ee}{\end{equation}}
\newcommand{\ba}{\begin{array}}
\newcommand{\ea}{\end{array}}
\newcommand{\bt}{\begin{tabular}}
\newcommand{\et}{\end{tabular}}
\newcommand{\un}{\underline}
\newcommand{\ov}{\overline}
\newcommand{\ds}{\displaystyle}

\newcommand{\ct}{\cite}

\newcommand{\lb}{\label}
\newcommand{\bp}{\begin{picture}}
\newcommand{\ep}{\end{picture}}
\newcommand{\bfi}{\begin{figure}}

\newcommand{\plaqr}{$\bigl(\;\;\raisebox{0pt}{\mbox{\framebox(0,7.6){\phantom{a}}\hspace{-4.2mm}
\dashbox{1.5}(8,8)[b]{\phantom{a}}}}\bigr)$}
\newcommand{\plaql}{$\bigl(\raisebox{0pt}{\mbox{\framebox(0,7.6){\phantom{a}}\hspace{-1.4mm}
\dashbox{1.5}(8,8)[b]{\phantom{a}}}}\bigr)$}
\newcommand{\plaqt}{$\bigl(\raisebox{0pt}{\mbox{\raisebox{8pt}{\framebox(7.6,0){\phantom{a}}}\hspace{-4.1mm}
\dashbox{1.5}(8,8)[b]{\phantom{a}}}}\bigr)$}
\newcommand{\plaqb}{$\bigl(
\raisebox{0pt}{\mbox{\framebox(7.6,0)
{\phantom{a}}\hspace{-4.1mm}
\dashbox{2}(8,8)[b]{\phantom{a}}}}\bigr)
$}

\newcommand{\link}{\begin{array}{l}\begin{picture}(22,4)
    \put(0,2.5){\circle*{4}}
    \put(20,2.5){\circle*{4}}
    \put(0,2.5){\line(1,0){20}}
\end{picture}\end{array}}

\begin{document}

\vspace{3cm}

\title{\Large\bf {Phase Transition in Gauge
Theories and the Planck Scale Physics.}}

\vspace{3cm}

\author{{\bf L.V.Laperashvili}
\footnote{{\bf E-mail}:laper@heron.itep.ru}\\
\it Institute of Theoretical and Experimental Physics,\\
\it B.Cheremushkinskaya 25, 117218 Moscow, Russia \\[0.2cm]
{\bf D.A.Ryzhikh}
\footnote{{\bf E-mail}:ryzhikh@heron.itep.ru}\\
\it Institute of Theoretical and Experimental Physics,\\
\it B.Cheremushkinskaya 25, 117218 Moscow, Russia}

\date{}

\maketitle

\begin{abstract}

The present paper is based on the modified part of the
review "Random Dynamics and Multiple Point Model"\: by
L.V.Laperashvili, H.B.Nielsen, D.A.Ryzhikh and N.Stillits,
in preparation for publication in Russian, which contains
the results of our joint activity with H.B.Nielsen
concerning the investigations of phase transitions in gauge
theories. In this review we have presented the main ideas
of the Nielsen's Random Dynamics (RD) and his achievements
(with co-authors) in the Anti-Grand Unification Theory
(AGUT) and Multiple Point Model (MPM). We have considered
also the theory of Scale Relativity (SR) by L.Nottale,
which has a lot in common with RD: both theories lead to
the discreteness of our space-time, giving rise to the new
description of physics at very small distances. In this
paper we have demonstrated the possibility of [SU(5)]$^3$
SUSY unification with superparticles of masses $M\approx
10^{18.3}$ GeV and calculated its critical point --
critical value of the inverse finestucture constant -- at
$\alpha_{5,crit}^{-1} = \alpha_5^{-1}(\mu_{Pl})\approx
34.0$ (close to $\alpha_{GUT}^{-1}\approx 34.4$) with a
hope that such an unified theory approaches the
(multi)critical point at the Planck scale.

\end{abstract}

\vspace*{-18cm}
\begin{flushright}
{\large Preprint ITEP~~24--01/~~~~}
\end{flushright}

\thispagestyle{empty}

\newpage
\vspace*{7mm}

\noindent
\hspace*{8.4cm}\parbox{68mm}{\itshape{
This review and new investigations are dedicated
to the 60th jubilee of the outstanding physicist of
Denmark {\itshape\bfseries Holger Bech Nielsen}, professor of the Niels Bohr Institute,
whose achievements are well--known in the World physics.}}\\

\section{Introduction: Random Dynamics}

The goal of physics always has been to describe an enormous phenomena
existing in Nature by a few number of rules called as "the laws of physics".
Trying to look insight the Nature and considering the physical processes
at small and very small distances, physicists have made attempts to
explain the well--known laws of physics as a consequence of the more
fundamental laws of Nature. The contemporary physics (high energy physics,
nuclear physics, solid state physics, astrophysics, cosmology) essentially
is based on the quantum theory and theory of general relativity. But on
the present level of our knowledge we strongly suspect that these theories
are not fundamental, they are a consequence of the more fundamental laws
of physics. On this way, Random Dynamics (RD)
was suggested and developed in Refs.[\citen{2a}--\citen{2m}] as a
theory of physical processes proceeding at small distances of order of the
Planck length $\lambda_P=M_{Pl}^{-1}$:
\begin{equation}
                M_{Pl}=1.22\cdot 10^{19}\,{\mbox{GeV}}.     \lb{1}
\end{equation}
RD tries to derive the laws of physics known today
by the use of almost no assumptions.

The modern physics of electroweak and strong interactions is described by
the Standard Model (SM) unifying the Glashow--Salam--Weinberg electroweak
theory and QCD, theory of strong interactions (see \ct{1a}).

The gauge group of SM is:
\begin{equation}
   SMG = SU(3)_c\times SU(2)_L\times U(1)_Y,            \lb{2}
\end{equation}
which describes the present elementary particle physics up to the
scale $\sim 100$ GeV.

SM contains 19 degrees of freedom: three independent gauge coupling
constants $g_i$ (i=1,2,3 correspond to the groups U(1), SU(2), SU(3));
three masses of leptons, six quark masses, the Higgs boson mass,
four quark mixing angles and two topological angles $\theta_{SU(2)}$
and $\theta_{QCD}$ from the topological terms in SU(2) and SU(3),
respectively.
A large part of these degrees of freedom are related to the coupling
of the Higgs sector to itself and to the matter sector. But at this
stage of our knowledge we have already had questions which have no
answers at present: Why has SM the Lie group $S(U(2)\times U(3))$?
Why have we three generations of the "fundamental" particles (at least,
within the energies of today experiment)? Do more generations exist?
And more fundamental question: Why quantum physics at small distances?

These questions lead to the conclusion that SM is not a TOE (Theory of
Everything). It looks like that SM is only a low-energy limit of
a more fundamental theory. The efforts to construct such a fundamental
theory have led to Grand Unified Theories (GUT), especially supersymmetric
GUT (SUSY GUT) which had an aim to give an unified selfconsistent
description of electroweak and strong interactions by one simple group
of symmetry SU(5), SU(6), SO(10), or $E_8$ (see reviews \ct{1r},\ct{2r}).
But at present time the experiment does not indicate any manifestation
of these theories.

The next step to search the fundamental theory was a set of
theories describing the extended objects: string, superstring and
M--theories. They came into existence due to the necessity
of unification of electroweak and strong interactions with gravity
[\citen{5r}--\citen{8r}].

RD is an alternative of these theories. The above mentioned theories
are based on some fixed axioms. If namely one of these axioms is
changed a little bit, the theory also is changed and often becomes not even
consistent. In RD we have the very opposite case.
RD is based on the very general assumptions which take place at
fundamental scale.

Wondering what fundamental laws of physics lead to the description of
the low--energy SM phenomena, observed by today experiment, we can consider
two possibilities:

\vspace{0.1cm}

1. A simple and fine theory underlies in physics: at very small
(Planck scale) distances
{\un{in our continuous space--time}}, there exists a theory with a high
symmetry, for example, such an extension of superstrings as M--theory
\ct{9a}, which unifies all known theories of superstrings (I, IIA, IIB,
SO(32), $E_8\times E_8$, etc.) with 11--dimensional supergravity.

\vspace{0.1cm}

2. The fundamental laws of Nature are so complicated that it is
preferable to think that at very small distances there are no laws
at all: {\un{our space--time is discrete}} and the physical processes
are described randomly.
Then the fundamental law of Nature is one, which is randomly chosen
from a large set of sufficiently complicated theories, and this one
leads to the laws observed in the low--energy limit by our experiment.

\vspace{0.1cm}

The item 2 is a base of RD theory.

\section{Theory of Scale Relativity}

The theory of scale relativity \ct{10na} is also related with item 2 of
Introduction and has a lot in common with RD. This is a theory representing
a new approach to understanding a quantum mechanics and various problems
of scales in today's physics. As it was mentioned in Introduction, the
contemporary physics is based on two main theories: the theory of relativity
(special and general, which include classical physics) and quantum mechanics
(developed into quantum field theories). Both theories are very effective
and precise in their predictions, but they are founded on completely
different grounds, even contradictory in appearance, and are described
by quite different mathematical apparatus. General relativity is based on the
fundamental physical principles, such as principles of general covariance
and equivalence. The mathematical description of this theory follows
from these principles. On the contrary, quantum mechanics is an axiomatic
theory (at least at present time), and its mathematical apparatus is
not understood on a more fundamental level yet. Such a situation shows
an apparent contradiction in physics: two opposite theories, classical and
quantum, cohabit there. In particular, gravity up to now has not
selfconsistent description in terms of the quantum field theory.
These and other signs indicate that physics is still in infancy: present
theory is not able to describe the two "tails" of the physical world --
very small and very large time and length scales.

It is clear now that the observed properties
of the quantum world cannot be reproduced by Riemannian geometry.
The space--time description cannot be based on particular fields, but
only on the universal properties of the matter. That is,
we have a necessity to introduce new concepts in physics.

The author of theory of scale relativity L.Nottale \ct{10na}, his
collaborators \ct{10nb} and B.G.Sidharth \ct{10nc}
assumed that the previous geometrical description of the quantum
properties of microphysics is impossible. They suggested to consider the
{\it non--differentiable space--time} and build the microphysical world
on the concept of the {\it fractal space--time}.  Moreover, this
non--differentiability implies {\it an explicit dependence of space--time on
scale}. The principle of relativity can be applied not only to motion, but also
to scale transformations, because the resolution of measurements must be taken
into account in definition of the coordinate system. The resolution of the
measurement apparatus plays in quantum physics a completely new role with
respect to the classical, since the result of measurements depends on it, as a
consequence of Heisenberg's relations.  Any set of physical data has its sense
only when it is accompanied by the measurement errors or uncertainties. More
generally, the resolution characterizes the system under consideration.
Complete information about the measurement of position and time can be
obtained when not only space--time coordinate (t,x,y,z), but also
resolutions ($\Delta t,\,\Delta x,\,\Delta y,\,\Delta z$) are given.
As a result, we expect space--time to be described by a metric element
based on generalized, explicitly scale--dependent, metric potentials:
\begin{equation}
g_{\mu\nu} = g_{\mu\nu}(t,x,y,z;\,\Delta t,\,\Delta x,\,\Delta y,\,\Delta z).
                                          \lb{3}
\end{equation}

\subsection{The concept of fractals}

It was shown in Refs.\ct{10nb} that a continuous but non--differentiable
space--time is necessarily fractal. Here a word {\it fractal} \ct{11n}
means an object or space showing structures at all scales, or on a wide
range of scales.

Let us consider a fractal "curve" in the ${\bf R}^2$ or ${\bf C}$ plane.
It can be obtained from an initial curve ${\bf F_1}$ made up of $p$
segments of equal length $1/q$ which connect the origin of the coordinate
system $XY$ to the point [0,1] (see Fig.1). If $\omega_{j+1}$ is the
polar angle of the $j$th segment and ${\bf Z}_j={\bf X}_j + i{\bf Y}_j=
(1/q){\bf \Sigma}_{k=o}^{j-1}e^{i\omega_k}$ is the complex coordinate
of a breaking point ${\bf P}_j$, then we have the following relations:
\begin{equation}
   \sum_{k=o}^{p-1}\,e^{i\omega_k}=q,
           \quad {\bf Z}_{j+1} - {\bf Z}_j = \frac{1}{q}e^{i\omega_j}.
                                                    \lb{4}
\end{equation}
A curve ${\bf F}_2$ can be built by substituting each segment of
${\bf F}_1$ by ${\bf F}_2$ itself, scaled at its length $1/q$, as
it is shown in Fig.2. Substitution of each segment of ${\bf F}_n$
by $q^{-n}{\bf F}_1$ giving ${\bf F}_{n+1}$ leads to the fractal
curve ${\bf F}$ which is a result of an infinite sequence of
these steps.

The curve {\bf F} can be parametrized by a real number $x\in [0,1]$
developed in the counting base $p$ in the form:
\begin{equation}
          x = 0.x_1x_2... = \sum_{k=1}^{\infty}x_kp^{-k}.   \lb{5}
\end{equation}
In this case we have so called "the Peano curve" \ct{11n}.

The fractal is completely defined when the complex coordinate ${\bf Z}(x)$
of the point on ${\bf F}$ parametrized by $x$ is known. According to
the above building process of ${\bf F}$, it is easy to obtain ${\bf Z}(x)$:
\begin{equation}
  {\bf Z}(x) ={\bf Z}_{x_1} + q^{-1}e^{i\omega_{x_1}}[{\bf Z}_{x_2} +
          q^{-1}e^{i\omega_{x_2}}[{\bf Z}_{x_3} + ...]],          \lb{6}
\end{equation}
and finally:
\begin{equation}
 {\bf Z}(x) = q\sum_{k=1}^{\infty}{\bf Z}_{x_k}e^{i\sum_{l=1}^{k-1}
                   {\omega_{x_l}}}q^{-k}.                  \lb{7}
\end{equation}
It is well--known \ct{11n} that a fractal dimension lies between 1 and 2,
and a fractal curve within a plane is intermediate between a line and
a surface. Indeed, while it may be built by adding segments, it may be
also be obtained by deleting surfaces \ct{11n}.

The fractal dimension is given by $D=\log\,p/\log\,q$. The length and
surface of the ${\bf F}_n$ are:
$$
   {\cal L}/{{\cal L}_0} = (p/q)^n = q^{n(D-1)},
$$
\begin{equation}
   {\cal S}/{{\cal S}_0} = (p/q^2)^n = q^{n(D-2)}.           \lb{8}
\end{equation}
We see that however small is the difference of parameters $x(2)-x(1)$
for two points $M_2$ and $M_1$ on the fractal, the distance in the plane
$|{\bf Z}(x_2) - {\bf Z}(x_1)|$ vanishes, then the fractal length goes
to infinity (but surface vanishes). From this "paradox" we conclude that
the description of the fractal curve ${\bf F}$ by the real coordinate
$x$ is unsufficient and another formalism is needed. But we don't consider
here the "non--standard analysis" of fractals, referring to the book \ct{11n}.

\subsection{Standard fractal scale-invariant laws}

Let us consider a non--differentiable coordinate system. The basic theorem
of the theory of scale relativity \ct{10nb} implies that the length of
the fractal curve $\cal L$ is an explicit function of the resolution
interval $\epsilon$: $\cal L = \cal L(\epsilon)$. At the first step
it was assumed that this function obeys the simplest possible (first order)
scale differential equation:
\begin{equation}
    \frac{d\ln{\cal L}}{d\ln(\lambda/\epsilon)} = \delta,   \lb{9}
\end{equation}
where $\delta$ is a constant. The solution of Eq.(\ref{9}) is a fractal
power--law dependence:
\begin{equation}
{\cal L}={{\cal L}_0}(\lambda/\epsilon)^{\delta},
\lb{10}
\end{equation}
here $\delta$ is the scale dimension: $\delta= D - D_T$, where $D$
and $D_T$ are the fractal and topological dimensions, respectively.

The group of scale transformations has the Galilean structure:
it transforms in a scale transformation $\epsilon\to \epsilon'$ as
\begin{equation}
 \ln\frac{{\cal L}(\epsilon')}{{\cal L}_0} =
 \ln\frac{{\cal L}(\epsilon)}{{\cal L}_0} +
 \delta(\epsilon)\ln\frac{\epsilon}{\epsilon'},\quad{\mbox{with}}
 \quad \delta(\epsilon')=\delta(\epsilon).              \lb{11}
\end{equation}
From Eq.(\ref{11}) the structure of the Galileo group is confirmed by
the law of composition of dilations $\epsilon\to \epsilon' \to \epsilon''$,
which leads to the relation $\ln\rho''=\ln\rho + \ln\rho'$, with
$\rho=\epsilon'/\epsilon$, $\rho'=\epsilon''/\epsilon'$ and $\rho''=
\epsilon''/\epsilon$.

\subsection{Breaking of scale symmetry}

One can assume also the existence of a first order differential
equation for the function ${\cal L}(\epsilon)$ which is similar to
the renormalization group equation (RGE):
\begin{equation}
           \frac{d{\cal L}}{d\ln{\epsilon}} =\beta({\cal L}).
                                                            \lb{12}
\end{equation}
Here a scale variable $\epsilon$ is a resolution.

The function $\beta({\cal L})$ is a priori unknown, but it is possible
to consider its Taylor expansion giving a perturbative series:
\begin{equation}
         \frac{d{\cal L}}{d\ln{\epsilon}} = a + b{\cal L} + ... \lb{13}
\end{equation}
The solution of this equation, with $\lambda$ as a constant of integration,
has the following form:
\begin{equation}
     {\cal L} = {\cal L}_0[1 + (\frac{\lambda}{\epsilon})^{\delta}],
                                                   \lb{14}
\end{equation}
where $\delta = - b$.

From Eq.(\ref{14}) we see for $\delta > 0$ a small--scale fractal
behavior of $\cal L$ which is broken at larger scales,
but for $\delta < 0$ this fractal behavior
takes place for large scales and is broken at smaller scales.

\subsection{Special scale relativity}

We know that the Galileo group of motion is a degeneration of the more
general Lorentz group. Assuming that the same is true for scale laws,
we can use a principle of scale relativity and consider instead of
the Galilean law of composition of dilations:
\begin{equation}
      \ln(\frac{\epsilon'}{\lambda}) =
            \ln{\rho} + \ln(\frac{\epsilon}{\lambda})     \lb{15}
\end{equation}
the more general Lorentzian law:
\begin{equation}
      \ln(\frac{\epsilon'}{\lambda}) = \frac
            {\ln{\rho} + \ln({\epsilon}/{\lambda})}
         {1 + \ln{\rho}\ln{(\epsilon/\lambda)}/\ln^2(\lambda_P/\lambda)}.
                                                     \lb{16}
\end{equation}
The scale dimension $\delta$ becomes itself a variable (see \ct{10nb}):
\begin{equation}
     \delta(\epsilon) = \frac{1}{\sqrt{1 - \ln^2(\epsilon/\lambda)/
                   \ln^2(\lambda_P/\lambda)}},             \lb{17}
\end{equation}
where $\lambda$ is the fractal--nonfractal transition scale.

\subsection{Fundamental scale}

In the law given by Eq.(\ref{17}) there exists a minimal scale of space--
time resolution $\epsilon_{min} = \lambda_P$, which is invariant under
dilations and contractions. It plays the same role for scales as the
velocity of light for motion in the Einstein's special theory of relativity.
This invariant length scale is the minimal, i.e. {\it fundamental scale}
of Nature. It is natural to identify it with the Planck scale, $\lambda_P =
{(\hbar G/c^3)}^{1/2}$. The Planck length and time scales thus appear as natural
units of length and time intervals.

If the fundamental scale exists in Nature, then our (3+1)--dimensional
space is discrete on the fundamental level. This hypothesis is a base
of RD theory by H.B.Nielsen \ct{2a}.  It is an initial point of view,
but not an approximation.

\section{Lattice Theories}

The lattice model of gauge theories is the most convenient formalism for the
realization of RD ideas.
In the simplest case we can imagine our space--time as a regular hypercubic
(3+1)--lattice with the parameter $a$ equal to the fundamental scale:
\begin{equation}
 a = \lambda_P = 1/M_{Pl}\sim 10^{-33}\,{\mbox {sm}}.    \lb{18a}
\end{equation}

\subsection{Mathematical structure of lattice gauge theories}

A lattice contains sites, links and plaquettes.
Link variables defined on the edges of the lattice are fundamental
variables of the lattice theory. These variables are simultaneously
the elements of gauge group $G$, describing a symmetry of the
corresponding lattice gauge theory:
\begin{equation}
{\cal U}(x\link y) \in G.
\lb{18}
\end{equation}
It is easy to understand the sense of this variable turning to the
differential geometry of the continuum space--time in which our gauge fields
${\hat A}_{\mu}(x)$ exist, where the quantity
\begin{equation}
      {\hat A}_{\mu}(x) = gA^j_{\mu}(x) t^j      \lb{24}
\end{equation}
contains the generator $t^j$ of the group $G$.
For $G=SU(3)$ we have $\,t^j = \lambda^j/2$,
where $\lambda^j$ are the well--known Gell--Mann matrices.

For $G=U(1)$:
\begin{equation}
      {\hat A}_{\mu}(x) = gA_{\mu}(x).      \lb{25}       
\end{equation}
Such a space geometrically is equivalent to curvilinear
space and an operator, which compares fields at different points,
is an operator of the parallel transport between the points $x,y$:
\begin{equation}
 {\cal U}(x,y) = Pe^{i\int_{C_{xy}}{\hat A}_{\mu}(x)dx^{\mu}}, \lb{19}
\end{equation}
where $P$ is the path ordering operator and $C_{xy}$ is a curve from
point $x$ till point $y$. Moreover, the operator:
\begin{equation}
    W = Tr(Pe^{i\oint_C{\hat A}_{\mu}(x)dx^{\mu}})  \lb{20}
\end{equation}
is the well--known Wilson loop.
In the case of scalar field  $\phi(x)$, interacting with gauge field
$A_{\mu}$, we have an additional gauge invariant observable:
\begin{equation}
     \phi^{+}(y)[Pe^{i\int_{C_{xy}}{\hat A}_{\mu}(x)dx^{\mu}}]\phi(x). \lb{21}
\end{equation}
The link variable (\ref{18}) is a lattice version of Eq.(\ref{19}):
\begin{equation}
   {\cal U}(x\link y) = e^{i\Theta_{\mu}(n)} \equiv {\cal U}_{\mu}(n).
                                \lb{22}
\end{equation}
This link variable connects the point $n$ and the point
$n + a_{\mu}$, where the index $\mu$ indicates the direction of a link
in the hypercubic lattice with parameter $a$. Considering the
infinitesimal increment of the operator (\ref{19}) in the continuum
limit, we have:
\begin{equation}
\Theta_{\mu}(n) = a{\hat A}_{\mu}(x).
\lb{23}
\end{equation}
Plaquette variables are not independent because they are products
of link variables:
\begin{equation}
{\cal U}_p\equiv {\cal U}(\square){\stackrel{def}{=}}
      {\cal U}\mbox{\plaqb} {\cal U}\mbox{\plaqr} {\cal U}\mbox{\plaqt}
      {\cal U}\mbox{\plaql}.              \lb{26}
\end{equation}

\subsection{Lattice actions}

The lattice action $S[{\cal U}]$ is invariant under the gauge transformations
on a lattice:
\bc
     ${\cal U} (x\link y) \longrightarrow \Lambda(x)
      {\cal U} (x\link y)\Lambda^{-1}(y),$
\ec
where $\Lambda(x)\in G$. The simplest action $S[{\cal U}]$ is given by the following expression:
\begin{equation}
     S[{\cal U}_p] = \sum_q \frac{\beta_q}{{\mbox{dim}}\,q}\sum_p
                  Re Tr {\cal U}_p^{(q)}.
                                                   \lb{28}
\end{equation}
Here $q$ is the index of representation of the group $G$,
${\mbox{dim}}\, q$
is the dimension of this representation, and $\beta_q =
1/g^2_q$, where $g_q$ is the coupling constant of gauge fields corresponding
to the representation $q$.

The path integral
\begin{equation}
          Z = \int D{\cal U}(\link)e^{ - S[{\cal
          U}(\link)]},
                                                     \lb{29}
\end{equation}
which is an analogue of the partition function, describes
the lattice gauge theory in the Euclidean four--dimensional space.  It is
necessary to construct the lattice field theory such that for $a \to 0$, i.e.
in the continuum limit, it leads to a regularized smooth gauge theory of fields
$A_{\mu}^j(x)$ (here $j$ is the symmetry subscript). In the opposite case, a
passage to the continuum limit is not unique \ct{12p}.

Let us consider the simplest case of the group $G = U(1)$, using the
only representation of this group in Eq.(\ref{28}):
\begin{equation}
           S[{\cal U}_p] = \beta \sum_p Re {\cal U}_p.           \lb{30}
\end{equation}
Here the quantity ${\cal U}_p$ is given by Eq.(\ref{26}) in
which the link variables ${\cal U}(\link)$ are complex numbers with their moduli
equal to unity, i.e.,
\begin{equation}
{\cal U}(x\link y) = \{z | z\in {\bf C},\, |z| = 1\}.   \lb{31}
\end{equation}
In the lattice model, the Lorentz gauge condition
has the form
\begin{equation}
\prod_{x\link y}{\cal U}(x\link y) = 1.
   \lb{32}
\end{equation}
Introducing the notation
\begin{equation}
                z = e^{i\Theta} ,     \lb{33}
\end{equation}
we can write:
\begin{equation}
     {\cal U}_p = e^{i\Theta_p}.           \lb{34}
\end{equation}
The variables ${\cal U}_p$ satisfy the identity:
\begin{equation}
      \prod_{\square \,\in
   ({\mbox{lattice\,\,cube}})}{\cal U}(\square) = I,   \lb{35}
\end{equation}
which is called the Bianchi identity. In Eq.(\ref{35}) the product is
taken over all plaquettes belonging to the cell (cube) of the hypercubic
lattice.

From Eqs.(\ref{30}) and (\ref{34}), the simplest lattice
$U(1)$ action has the form:
\begin{equation}
        S[{\cal U}_p] = \beta \sum_p \cos {\Theta}_p.    \lb{35a}
\end{equation}
For the compact lattice QED: $\beta=1/e_0^2$, where $e_0$ is the bare
electric charge.

The lattice $SU(N)$ gauge theories were first introduced by K.Wilson \ct{1s}
for studying the problem of confinement. He suggested the following
simplest action:
\begin{equation}
      S = - \frac{\beta}{N}\sum_p Re(Tr\,{\cal U}_p),          \lb{36}
\end{equation}
where the sum runs over all plaquettes of a hypercubic lattice and
${\cal U}_p\equiv {\cal U}(\square)$ belongs to the fundamental
representation of $SU(N)$.

Monte Carlo simulations of these simple Wilson lattice
theories in four dimensions showed a (or an almost) second--order
deconfining phase transition for U(1) \ct{2s},\ct{3s}, a crossover
behavior for SU(2) and SU(3) \ct{4s},\ct{5s}, and a first--order
phase transition for SU(N) with $N\ge 4$ \ct{6s}. Bhanot and
Creutz \ct{7s},\ct{8s} have generalized the simple Wilson theory,
introducing two parameters in the $SU(N)$ action:
\begin{equation}
   S = \sum_p[-\frac{\beta_f}{N}Re(Tr\,{\cal U}_p) -
               \frac{\beta_A}{N^2-1}Re(Tr_A{\cal U}_p)],   \lb{37}
\end{equation}
where $\beta_f$, $Tr$ and $\beta_A$, $Tr_A$ are
respectively the lattice constants and traces in the fundamental and
adjoint representations of $SU(N)$.
The phase diagrams obtained for the generalized lattice SU(2) and
SU(3) theories (\ref{37}) by Monte Carlo methods in Refs.\ct{7s},\ct{8s}
(see also \ct{8sa}) are shown in Fig.3(a,b). They indicated
the existence of a triple point which is a boundary point of three
first--order phase transitions: the "Coulomb--like" and confining
$SU(N)/Z_N$, $Z_N$ phases meet together at this point. From the triple
point emanate three phase border lines which separate the corresponding
phases. The $Z_N$ phase transition is a "discreteness" transition, occurring
when lattice plaquettes jump from the identity to nearby elements
in the group. The $SU(N)/Z_N$ phase transition is due to a condensation
of monopoles (a consequence of the non-trivial $\Pi_1$ of the group).

Monte Carlo simulations of the U(1) gauge theory described by the
two-parameter lattice action \ct{9s},\ct{10s}:
\begin{equation}
     S = \sum_p[\beta^{lat} \cos \Theta_p + \gamma^{lat} \cos2\Theta_p],\quad
     {\mbox {where}} \quad {\cal U}_p = e^{i\Theta_p},     \lb{38}
\end{equation}
also indicate the existence of a triple point on the corresponding
phase diagram: "Coulomb--like", totally confining and $Z_2$ confining
phases come together at this triple point (see Fig.4).

Lattice theories are given in reviews \ct{12p}.
The next efforts of the lattice simulations for the SU(N) gauge theories
are presented in the review \ct{10as}.

\subsection{Lattice artifact monopoles}

Lattice monopoles are responsible for the confinement in lattice gauge
theories what is confirmed by many numerical and theoretical investigations
(see reviews \ct{11s} and papers \ct{12s}).

In the compact lattice gauge theory the monopoles are not physical objects:
they are lattice artifacts driven to infinite mass in the continuum
limit. Weak coupling ("Coulomb") phase terminates because of the
appearance for non-trivial topological configurations which are able
to change the vacuum. These topological excitations are closed monopole
loops (or universe lines of the monopole-antimonopole pairs). When these
monopole loops are long and numerous, they are responsible for the
confinement. But when they are dilute and small, the Coulomb ("free photon")
phase appears. Banks et al.\ct{[22]} have shown that in the Villain
form of the $U(1)$ lattice gauge theory \ct{[23]} it is easy to
exhibit explicitly the contribution of the topological excitations.  The
Villain lattice action is:
\begin{equation}
  S_V =  (\beta/2))\sum_p {(\Theta_p - 2\pi k)}^2, \qquad k\in Z.
                                                \lb{39}
\end{equation}
In such a model the partition function $Z$ may be written in a factorized form:
\begin{equation}
                Z = Z_C Z_M,                     \lb{40}     
\end{equation}
where $Z_C$ is a part describing
the photons:
\begin{equation}
        Z_C \sim \int\limits_{-\infty}^{\infty}d\Theta
      \exp [-\frac{\beta}{2} \sum_{\square}{\Theta}^2(\square)],  \lb{41}  
\end{equation}
and $Z_M$ is the partition function of a gas of the monopoles \ct{[24]},
\cite{[25]}:
\begin{equation}
  Z_M \sim \sum_{m\in Z} \exp [-2{\pi}^2\beta \sum_{x,y}m(x)v(x-y)m(y)].
                                         \lb{42}           
\end{equation}

In Eq.(\ref{42}) $v(x-y)$ is a lattice version of $1/r$-potential and
$m(x)$ is the charge of the monopole sitting in an elementary cube $c$
of the dual lattice, which can be simply expressed in terms of the
integer variables $n_p$ :
\begin{equation}
             m =\sum_{p\in \partial c}n_p,
                                            \lb{43}    
\end{equation}
where $n_p$ is a number of Dirac strings passing through the plaquettes of
the cube $c$.

The Gaussian part $Z_C$ provides the usual Coulomb potential, while
the monopole part $Z_M$ leads, at large separations, to a linearly confining
potential.

It is more complicated to exhibit the contribution of monopoles even in the
$U(1)$ lattice gauge theory described by the simple Wilson action (\ref{35a}).
Let us consider the Wilson loop as a rectangle of length $T$ in the
1-direction (time) and width $R$ in the 2-direction (spacelike distance),
then we can extract the potential $V(R)$ between two static charges of
opposite signs:
\begin{equation}
         V(R) = - \lim_{T\to\infty}\frac{1}{T}\log <W>,   \lb{44}  
\end{equation}
and obtain:
\begin{equation}
   V(R) = - \frac{\alpha (\beta)}{R}\quad -
        {\mbox {in\,the\,"Coulomb"\,phase}},                     \lb{45}
\end{equation}
\begin{equation}
  V(R) = \sigma R -\frac{\alpha (\beta)}{R} + O(\frac{1}{R^3}) +
  const \, - {\mbox {in\,the\,confinement\,phase}}.  \lb{46}
\end{equation}

\subsection{The behavior of electric fine structure constant $\alpha$
near the phase transition point. "Freezing" of $\alpha$}

The lattice investigators were not able to obtain the lattice
triple point values of $\alpha_{i,crit}$ by Monte Carlo simulations
method. Only the critical value of the effective electric fine structure
constant $\alpha$
was obtained in Ref.\ct{10s} in the compact QED described by the Wilson and
Villain actions (\ref{35a}) and (\ref{39}), respectively:
\begin{equation}
\alpha_{crit}^{lat}\approx 0.20\pm 0.015\quad
{\mbox{and}} \quad {\tilde \alpha}_{crit}^{lat}\approx 1.25\pm 0.10
\quad{\mbox{at}}\quad
\beta_T\equiv\beta_{crit}\approx{1.011}.
                                                     \lb{47}
\end{equation}
Here:
\begin{equation}
    \alpha = \frac{e^2}{4\pi}\quad{\mbox{and}}\quad
    \tilde \alpha = \frac{g^2}{4\pi}                   \lb{47*}
\end{equation}
are the electric and magnetic fine structure constants, containing
the electric charge $e$ and magnetic charge $g$.

The result of Ref.\ct{10s} for the behavior of $\alpha(\beta)$ in the vicinity
of the phase transition point $\beta_T$ is shown in Fig.5(a) for the Wilson
and Villain lattice actions. Fig.5(b) demonstrates the comparison of the
function $\alpha(\beta)$ obtained by Monte Carlo method for the Wilson
lattice action and by theoretical calculation of the same quantity.
The theoretical (dashed) curve was calculated by so-called "Parisi improvement
formula" \ct{13p}:
\begin{equation}
    \alpha (\beta )=[4\pi \beta W_p]^{-1}.     \lb{48}
\end{equation}
Here $W_p=<\cos \Theta_p >$ is a mean value of the plaquette energy.
The corresponding values of $W_p$ are taken from Ref.\ct{9s}.

The theoretical value of $\alpha_{crit}$ is less than the "experimental"
(Monte Carlo) value (\ref{47}):
\begin{equation}
      \alpha_{crit}\mbox{(in\,\,lattice\,\,theory)}\approx{0.12}.
                                                    \lb{49}
\end{equation}
This discrepancy between the theoretical and "experimental" results
is described by monopole contributions: the fine structure constant is
renormalized by an amount proportional to the susceptibility of the
monopole gas \cite{[24]}:
\begin{equation}
K = \frac{{\alpha}_{crit}\mbox{(Monte\,Carlo)}}{{\alpha}_
{crit}\mbox{(lattice\,theory)}}
\approx{\frac{0.20}{0.12}}\approx{1.66}.
\lb{50}
\end{equation}
Such an enhancement of the  critical fine structure constant is due to vacuum
monopole loops \ct{[25]}.

According to Fig.5(c):
\begin{equation}
   \alpha_{crit.,theor.}^{-1}\approx 8.                   \lb{50a}
\end{equation}
This result does not coincide with the lattice result (\ref{47}),
which gives the following value:
\begin{equation}
   \alpha_{crit.,lat.}^{-1}\approx 5.                   \lb{50b}
\end{equation}
The deviation of theoretical calculations of $\alpha(\beta )$ from the
lattice ones, which is shown in Fig.5(b,c),
has the following explanation: "Parisi improvement formula" (\ref{48})
is valid in Coulomb phase where the mass of artifact monopoles is infinitely
large and photon is massless. But in the vicinity of the phase
transition (critical) point the monopole mass $m\to 0$ and photon
acquires the non--zero mass $m_0\neq 0$ on the confinement side. This phenomenon leads to
the "freezing" of $\alpha$: the effective electric fine structure constant
is almost unchanged in the confinement phase and approaches to its maximal
value $\alpha=\alpha_{max}$. The authors of Ref.\ct{14p} predicted that
in the confinement phase, where we have the formation of strings, the fine
structure constant $\alpha$ cannot be infinitely large, but has the maximal
value:
\begin{equation}
\alpha_{max} \approx \frac{\pi}{12}\approx 0.26,             \lb{51}
\end{equation}
due to the Casimir effect for strings. The authors of Ref.\ct{14pa}
developed this viewpoint in spinor QED: the vacuum polarization
induced by thin "strings"--vortices of magnetic flux leads to the
suggestion of an analogue of the "spaghetti vacuum" \ct{14pb} as a possible
mechanism for avoiding the divergences of perturbative QED. According to
Ref.\ct{14pa}, the non--perturbative sector of QED arrests the growth
of the effective $\alpha$ to infinity and confirms the existence of
$\alpha_{max}$. This phenomenon was called "the freezing of
$\alpha$".

We see that Fig.5(a) demonstrates the tendency to freezing of
$\alpha$ in the vicinity of the phase transition point
$\beta = \beta_T$ for the compact QED.

The analogous phenomenon of the "freezing" of $\alpha_s$ was considered
in QCD in Refs.\ct{15p}.

\section{Higgs Monopole Model and Phase Transition in the
Regularized U(1) Gauge Theory}

The simplest effective dynamics describing the
confinement mechanism in the pure gauge lattice U(1) theory
is the dual Abelian Higgs model of scalar monopoles \ct{13s}
(see also Refs.\ct{11s} and \ct{12s}).

In the previous papers \ct{16p}-\ct{19p} the calculations of the U(1)
phase transition (critical) coupling constant were connected with the
existence of artifact monopoles in the lattice gauge theory and also
in the Wilson loop action model \ct{19p}.

In Ref.\ct{19p} we (L.V.L. and H.B.Nielsen) have put forward the speculations
of Refs.\ct{16p}-\ct{18p} suggesting that the modifications of the form of
the lattice action might not change too much the phase transition value of the
effective continuum coupling constant. The purpose was to investigate this
approximate stability of the critical coupling with respect to a somewhat
new regularization being used instead of the lattice, rather than just
modifying the lattice in various ways.
In \ct{19p} the Wilson loop action was considered in the
approximation of circular loops of radii $R\ge a$. It was shown that the
phase transition coupling constant is indeed approximately independent
of the regularization method: ${\alpha}_{crit}\approx{0.204}$,
in correspondence with the Monte Carlo simulation result on lattice:
${\alpha}_{crit}\approx{0.20\pm 0.015}$ (see Eq.(\ref{47}) and \ct{10s}).

But in Refs.\ct{20p}-\ct{23p}, instead of using the lattice or Wilson loop
cut--off, we have considered the Higgs Monopole Model (HMM) approximating
the lattice artifact monopoles as fundamental pointlike particles described
by the Higgs scalar field. Considering the renormalization group improvement
of the effective Coleman--Weinberg potential \ct{20s},\ct{21s}, written
in Ref.\ct{22p} for the dual sector of scalar electrodynamics in the
two--loop approximation, we have calculated the U(1) critical values of
the magnetic fine structure constant
${\tilde\alpha}_{crit} = g^2_{crit}/4\pi\approx 1.20$
and electric fine structure constant
$\alpha_{crit} = \pi/g^2_{crit}\approx 0.208 $ (by the Dirac relation).
These values coincide with the lattice result (\ref{47}).
The next subsections follow the review \ct{23p} of the HMM
calculations of the U(1) critical couplings.

\subsection{The Coleman-Weinberg effective potential in HMM}

As it was mentioned above, the dual Abelian Higgs model of scalar
monopoles (shortly HMM) describes the dynamics of confinement in lattice
theories. This model, first suggested in Refs.\ct{13s}, considers the
following Lagrangian:
\begin{equation}
    L = - \frac{1}{4g^2} F_{\mu\nu}^2(B) + \frac{1}{2} |(\partial_{\mu} -
           iB_{\mu})\Phi|^2 - U(\Phi),\quad              \lb{5y}
{\mbox{where}}\quad
 U(\Phi) = \frac{1}{2}\mu^2 {|\Phi|}^2 + \frac{\lambda}{4}{|\Phi|}^4
\end{equation}
is the Higgs potential of scalar monopoles with magnetic charge $g$, and
$B_{\mu}$ is the dual gauge (photon) field interacting with the scalar
monopole field $\Phi$.  In this model $\lambda$ is the self--interaction
constant of scalar fields, and the mass parameter $\mu^2$ is negative.
In Eq.(\ref{5y}) the complex scalar field $\Phi$ contains
the Higgs ($\phi$) and Goldstone ($\chi$) boson fields:
\begin{equation}
          \Phi = \phi + i\chi.             \lb{7y}
\end{equation}
The effective potential in the Higgs Scalar ElectroDynamics (HSED)
was first calculated by Coleman and Weinberg \ct{20s} in the one--loop
approximation. The general method of its calculation is given in the
review \ct{21s}. Using this method, we can construct the effective potential
for HMM. In this case the total field system of the gauge ($B_{\mu}$)
and magnetically charged ($\Phi$) fields is described by
the partition function which has the following form in Euclidean space:
\begin{equation}
      Z = \int [DB][D\Phi][D\Phi^{+}]\,e^{-S},     \lb{8y}
\end{equation}
where the action $S = \int d^4x L(x) + S_{gf}$ contains the Lagrangian
(\ref{5y}) written in the Euclidean space and gauge fixing action $S_{gf}$.

Let us consider now a shift:
\begin{equation}
 \Phi (x) = \Phi_b + {\hat \Phi}(x),                \lb{9y}
\end{equation}
with $\Phi_b$ as a background field, and calculate the
following expression for the partition function in the one-loop
approximation:
$$
  Z = \int [DB][D\hat \Phi][D{\hat \Phi}^{+}]
   \exp\{ - S(B,\Phi_b)
   - \int d^4x [\frac{\delta S(\Phi)}{\delta \Phi(x)}|_{\Phi=
   \Phi_b}{\hat \Phi}(x) + h.c. ]\}\\
$$
\begin{equation}
    =\exp\{ - F(\Phi_b, g^2, \mu^2, \lambda)\}.      \lb{10y}
\end{equation}
Using the representation (\ref{7y}), we obtain the effective potential:
\begin{equation}
  V_{eff} = F(\phi_b, g^2, \mu^2, \lambda)        \lb{11y}
\end{equation}
given by the function $F$ of Eq.(\ref{10y}) for the real constant background
field $ \Phi_b = \phi_b = \mbox{const}$. In this case the one--loop
effective potential for monopoles coincides with the expression of the
effective potential calculated by the authors of Ref.\ct{20s} for scalar
electrodynamics and extended to the massive theory (see review \ct{21s}):
$$
 V_{eff}(\phi_b^2) = \frac{\mu^2}{2} {\phi_b}^2 +
                 \frac{\lambda}{4} {\phi_b}^4
  + \frac{1}{64\pi^2}[ 3g^4 {\phi_b}^4\log\frac{\phi_b^2}{M^2} +
$$
\begin{equation}
   {(\mu^2 + 3\lambda {\phi_b}^2)}^2\log\frac{\mu^2 + 3\lambda\phi_b^2}{M^2}
 + {(\mu^2 +\lambda \phi_b^2)}^2\log\frac{\mu^2
 + \lambda \phi_b^2}{M^2}] + C,
                           \lb{12y}
\end{equation}
where $M$ is the cut--off scale and C is a constant not depending on
$\phi_b^2$.

The effective potential (\ref{11y}) has several minima. Their position
depends on $g^2, \mu^2$ and $\lambda$.
If the first local minimum occurs at $\phi_b=0$ and $V_{eff}(0)=0$,
it corresponds to the so-called "symmetrical phase", which is
the Coulomb-like phase in our description. Then it is easy to determine
the constant C in Eq.(\ref{12y}):
\begin{equation}
        C = - \frac{\mu^4}{16{\pi^2}}\log \frac {\mu}M,  \lb{13y}
\end{equation}
and we have the effective potential for HMM described
by the following expression:
\begin{equation}
V_{eff}(\phi^2_b)
= \frac{\mu^2_{run}}{2}\phi_b^2 + \frac{\lambda_{run}}{4}\phi_b^4
     + \frac{\mu^4}{64\pi^2}\log\frac{(\mu^2 + 3\lambda \phi_b^2)(\mu^2 +
       \lambda \phi_b^2)}{\mu^4}.
                                                \lb{14y}
\end{equation}
Here $\lambda_{run}$ is the running self--interaction constant
given by the expression standing in front of $\phi_b^4$ in Eq.(\ref{12y}):
\begin{equation}
  \lambda_{run}(\phi_b^2)
   = \lambda + \frac{1}{16\pi^2} [ 3g^4\log \frac{\phi_b^2}{M^2}
   + 9{\lambda}^2\log\frac{\mu^2 + 3\lambda \phi_b^2}{M^2} +
     {\lambda}^2\log\frac{\mu^2 + \lambda\phi_b^2}{M^2}].   \lb{15y}
\end{equation}
The running squared mass of the Higgs scalar monopoles also follows from
Eq.(\ref{12y}):
\begin{equation}
   \mu^2_{run}(\phi_b^2)
   = \mu^2 + \frac{\lambda\mu^2}{16\pi^2}[ 3\log\frac{\mu^2 +
   3\lambda \phi_b^2}{M^2} + \log\frac{\mu^2 + \lambda\phi_b^2}{M^2}].
                                  \lb{16y}
\end{equation}
As it was shown in Ref.\ct{20s}, the effective potential
can be improved by consideration of the renormalization
group equation (RGE).

\subsection{Renormalization group equations in HMM}

The RGE for the effective potential means that the potential cannot
depend on a change in the arbitrary parameter -- renormalization scale $M$:
\begin{equation}
         \frac {dV_{eff}}{dM} = 0.             \lb{17y}
\end{equation}
The effects of changing it are absorbed into
changes in the coupling constants, masses and fields, giving so--called
running quantities.

Considering the RG improvement of the effective potential \ct{20s}, \ct{21s}
and choosing the evolution variable as
\begin{equation}
                   t = \log(\phi^2/M^2),     \lb{18y}
\end{equation}
we have the following RGE for the improved $V_{eff}(\phi^2)$
with $\phi^2\equiv \phi^2_b$ \ct{22s}:
\begin{equation}
   (M^2\frac{\partial}{\partial M^2} + \beta_{\lambda}\frac{\partial}
{\partial\lambda} + \beta_g\frac{\partial}{\partial g^2} +
\beta_{(\mu^2)}\mu^2\frac{\partial}{\partial \mu^2} - \gamma\phi^2
\frac{\partial}{\partial \phi^2})V_{eff}(\phi^2) = 0,    \lb{19y}
\end{equation}
where $\gamma$ is the anomalous dimension and $\beta_{(\mu^2)}$,
$\beta_{\lambda}$ and $\beta_g$ are the RG $\beta$--functions for mass,
scalar and gauge couplings, respectively. RGE (\ref{19y}) leads to the
following form of the improved effective potential \ct{20s}:
\begin{equation}
     V_{eff} = \frac{1}{2}\mu^2_{run}(t)G^2(t)\phi^2 +
                 \frac{1}{4}\lambda_{run}(t)G^4(t)\phi^4.  \lb{20y}
\end{equation}
In our case:
\begin{equation}
 G(t) = \exp[-\frac{1}{2}\int_0^t dt'\,\gamma\left(g_{run}(t'),
         \lambda_{run}(t')\right)].                         \lb{21y}
\end{equation}
A set of ordinary differential equations (RGE) corresponds to Eq.(\ref{19y}):
\begin{equation}
    \frac{d\lambda_{run}}{dt} = \beta_{\lambda}\left(g_{run}(t),\,
                    \lambda_{run}(t)\right),      \lb{22y}
\end{equation}
\begin{equation}
    \frac{d\mu^2_{run}}{dt} = \mu^2_{run}(t)\beta_{(\mu^2)}
             \left(g_{run}(t),\,\lambda_{run}(t)\right),
                                                  \lb{23y}
\end{equation}
\begin{equation}
    \frac{dg^2_{run}}{dt} = \beta_g\left(g_{run}(t),\,\lambda_{run}(t)\right).
                                           \lb{24y}
\end{equation}
So far as the mathematical structure of HMM is equivalent
to HSED, we can use all results of the scalar electrodynamics
in our calculations, replacing the electric charge $e$ and photon
field $A_{\mu}$ by magnetic charge $g$ and dual gauge field $B_{\mu}$.

Let us write now the one--loop potential (\ref{14y}) as
\begin{equation}
         V_{eff} = V_0 + V_1,          \lb{25y}
\end{equation}
where
$$
   V_0 = \frac{\mu^2}2 \phi^2 + \frac{\lambda}4 \phi^4,\quad
  V_1 = \frac{1}{64\pi^2}[ 3g^4 {\phi}^4\log\frac{\phi^2}{M^2}
+ {(\mu^2 + 3\lambda {\phi}^2)}^2\log\frac{\mu^2 + 3\lambda\phi^2}{M^2}
$$
\begin{equation}
    + {(\mu^2 +\lambda \phi^2)}^2\log\frac{\mu^2
     + \lambda \phi^2}{M^2} - 2\mu^4\log\frac{\mu^2}{M^2}].
                                                    \lb{27y}
\end{equation}
We can plug this $V_{eff}$ into RGE (\ref{19y}) and obtain the
following equation (see \ct{21s}):
\begin{equation}
   ( \beta_{\lambda}\frac{\partial}{\partial \lambda} +
    \beta_{(\mu^2)}{\mu^2}\frac{\partial}{\partial \mu^2} -
    \gamma \phi^2 \frac{\partial}{\partial \phi^2}) V_0 =
     - M^2\frac{\partial V_1}{\partial M^2}.
                                           \lb{28y}
\end{equation}
Equating $\phi^2$ and $\phi^4$ coefficients, we obtain the expressions
of $\beta_{\lambda}$ and $\beta_{(\mu^2)}$ in the one--loop approximation:
\begin{equation}
    \beta_{\lambda}^{(1)}
= 2\gamma \lambda_{run}
           + \frac{5\lambda_{run}^2}{8\pi^2} +
                \frac{3g_{run}^4}{16\pi^2},         \lb{29y}
\end{equation}
\begin{equation}
    \beta_{(\mu^2)}^{(1)}
= \gamma + \frac{\lambda_{run}}{4\pi^2}. \lb{30y}
\end{equation}
The one--loop result for $\gamma$ is given in Ref.\ct{20s} for scalar field
with electric charge $e$, but it is easy to rewrite this $\gamma$--expression
for monopoles with charge $g=g_{run}$:
\begin{equation}
          \gamma^{(1)} = - \frac{3g_{run}^2}{16\pi^2}.   \lb{30ay}
\end{equation}
Finally we have:
\begin{equation}
\frac{d\lambda_{run}}{dt}\approx \beta_{\lambda}^{(1)} = \frac 1{16\pi^2}
( 3g^4_{run} +10 \lambda^2_{run} - 6\lambda_{run}g^2_{run}),
                                     \lb{31y}
\end{equation}
\begin{equation}
\frac{d\mu^2_{run}}{dt}\approx \beta_{(\mu^2)}^{(1)}
= \frac{\mu^2_{run}}{16\pi^2}( 4\lambda_{run} - 3g^2_{run} ).
                                                \lb{32y}
\end{equation}
The expression of $\beta_g$--function in the one--loop approximation
also is given by the results of Ref.\ct{20s}:
\begin{equation}
    \frac{dg^2_{run}}{dt}\approx
     \beta_g^{(1)} = \frac{g^4_{run}}{48\pi^2}.  \lb{33y}
\end{equation}

The RG $\beta$--functions for different renormalizable gauge theories with
semisimple group have been calculated in the two--loop approximation
\ct{22y}-\ct{27y} and even beyond \ct{28y}. But in this paper we made use the
results of Refs.\ct{22y} and \ct{25y} for calculation of $\beta$--functions
and anomalous dimension in the two--loop approximation, applied to the
HMM with scalar monopole fields. The higher approximations essentially
depend on the renormalization scheme \ct{28y}.
Thus, on the level of two--loop approximation we have for all
$\beta$--functions:
\begin{equation}
  \beta = \beta^{(1)} + \beta^{(2)},           \lb{34y}
\end{equation}
where
\begin{equation}
  \beta_{\lambda}^{(2)} = \frac{1}{(16\pi^2)^2}( - 25\lambda^3 +
   \frac{15}{2}g^2{\lambda}^2 - \frac{229}{12}g^4\lambda - \frac{59}{6}g^6),
                                                    \lb{35y}
\end{equation}
and
\begin{equation}
\beta_{(\mu^2)}^{(2)} = \frac{1}{(16\pi^2)^2}(\frac{31}{12}g^4 + 3\lambda^2).
                                             \lb{36y}
\end{equation}
The gauge coupling $\beta_g^{(2)}$--function is given by Ref.\ct{22y}:
\begin{equation}
     \beta_g^{(2)} = \frac{g^6}{(16\pi^2)^2}.  \lb{37y}
\end{equation}
Anomalous dimension follows from calculations made in Ref.\ct{25y}:
\begin{equation}
    \gamma^{(2)} = \frac{1}{(16\pi^2)^2}\frac{31}{12}g^4.
                                                   \lb{38y}
\end{equation}
In Eqs.(\ref{34y})--(\ref{38y}) and below, for simplicity, we have used the
following notations: $\lambda\equiv \lambda_{run}$, $g\equiv g_{run}$ and
$\mu\equiv \mu_{run}$.

\subsection{The phase diagram in HMM}

Now we  want to apply the effective potential calculation as a
technique for the getting phase diagram information for the condensation
of monopoles in HMM.
As it was mentioned in the subsection 4.1,
the effective potential (\ref{20y}) can have several minima. Their positions
depend on $g^2$, $\mu^2$ and $\lambda$. If the first local minimum occurs
at $\phi = 0$ and $V_{eff}(0) = 0$, it corresponds to the Coulomb--like phase.
In the case when the effective potential has the second local minimum at
$\phi = \phi_{min} \neq 0\,$ with $\,V_{eff}^{min}(\phi_{min}^2) < 0$,
we have the confinement phase. The phase transition between the
Coulomb--like and confinement phases is given by the condition when
the first local minimum at $\phi = 0$ is degenerate with the second minimum
at $\phi = \phi_0$.
These degenerate minima are shown in Fig.6 by the curve 1. They correspond
to the different vacua arising in this model. And the dashed curve 2
describes the appearance of two minima corresponding to the confinement
phases (see details in the next Section).

The conditions of the existence of degenerate vacua are given by the
following equations:
\begin{equation}
           V_{eff}(0) = V_{eff}(\phi_0^2) = 0,     \lb{39y}
\end{equation}
\begin{equation}
    \frac{\partial V_{eff}}{\partial \phi}|_{\phi=0} =
    \frac{\partial V_{eff}}{\partial \phi}|_{\phi=\phi_0} = 0,
\quad{\mbox{or}}\quad V'_{eff}(\phi_0^2)\equiv
\frac{\partial V_{eff}}{\partial \phi^2}|_{\phi=\phi_0} = 0,
                                                    \lb{40y}
\end{equation}
and inequalities
\begin{equation}
    \frac{\partial^2 V_{eff}}{\partial \phi^2}|_{\phi=0} > 0, \qquad
    \frac{\partial^2 V_{eff}}{\partial \phi^2}|_{\phi=\phi_0} > 0.
                                               \lb{41y}
\end{equation}
The first equation (\ref{39y}) applied to Eq.(\ref{20y}) gives:
\begin{equation}
    \mu^2_{run} = - \frac{1}{2} \lambda_{run}(t_0)\,\phi_0^2\, G^2(t_0),
\quad{\mbox{where}}\quad t_0 = \log(\phi_0^2/M^2).
                                    \lb{42y}
\end{equation}
Calculating the first derivative of $V_{eff}$ given by Eq.(\ref{40y}),
we obtain the following expression:
$$
      V'_{eff}(\phi^2) = \frac{V_{eff}(\phi^2)}{\phi^2}(1 +
 2\frac{d\log G}{dt}) + \frac 12 \frac{d\mu^2_{run}}{dt} G^2(t)
$$
\begin{equation}
     + \frac 14 \biggl(\lambda_{run}(t) + \frac{d\lambda_{run}}{dt} +
      2\lambda_{run}\frac{d\log G}{dt}\biggr)G^4(t)\phi^2.
                                                \lb{45y}
\end{equation}
From Eq.(\ref{21y}), we have:
\begin{equation}
          \frac{d\log G}{dt} = - \frac{1}{2}\gamma .  \lb{46y}
\end{equation}
It is easy to find the joint solution of equations
\begin{equation}
      V_{eff}(\phi_0^2) = V'_{eff}(\phi_0^2) = 0.       \lb{47y}
\end{equation}
Using RGE (\ref{22y}), (\ref{23y}) and Eqs.(\ref{42y})--(\ref{46y}),
we obtain:
\begin{equation}
 V'_{eff}(\phi_0^2) =\frac{1}{4}( - \lambda_{run}\beta_{(\mu^2)} +
\lambda_{run} + \beta_{\lambda} - \gamma \lambda_{run})G^4(t_0)\phi_0^2 = 0,
                                                    \lb{48y}
\end{equation}
or
\begin{equation}
    \beta_{\lambda} + \lambda_{run}(1 - \gamma - \beta_{(\mu^2)}) = 0.
                                            \lb{49y}
\end{equation}
Substituting in Eq.(\ref{49y}) the functions
$\beta_{\lambda}^{(1)},\,\beta_{(\mu^2)}^{(1)}$ and $\gamma^{(1)}$
given by \mbox{Eqs.(\ref{29y})--(\ref{30ay})}, we obtain in the one--loop
approximation the following equation for the phase transition border:
\begin{equation}
     g^4_{PT} = - 2\lambda_{run}(\frac{8\pi^2}3 + \lambda_{run}).
                                                 \lb{50y}
\end{equation}
The curve (\ref{50y}) is represented on the phase diagram
$(\lambda_{run}; g^2_{run})$ of Fig.7 by the curve "1" which describes
the border between the Coulomb--like phase with $V_{eff} \ge 0$
and the confinement one with $V_{eff}^{min} < 0$. This border corresponds to
the one--loop approximation.

Using Eqs.(\ref{30ay})-(\ref{38y}), we are able to construct
the phase transition border in the two--loop approximation.
Substituting these equations into Eq.(\ref{49y}), we obtain the following
phase transition border curve equation in the two--loop approximation:
\begin{equation}
 3y^2 - 16\pi^2 + 6x^2 + \frac{1}{16\pi^2}(28x^3 + \frac{15}{2}x^2y +
  \frac{97}{4}xy^2 - \frac{59}{6}y^3) = 0,            \lb{51y}
\end{equation}
where $x = - \lambda_{PT}$ and $y = g^2_{PT}$ are the phase transition
values of $ - \lambda_{run}$ and $g^2_{run}$.
Choosing the physical branch corresponding to $g^2 \ge 0$ and $g^2\to 0$,
when $\lambda \to 0$, we have received curve 2 on the phase diagram
$(\lambda_{run}; g^2_{run})$ shown in Fig.7. This curve
corresponds to the two--loop approximation and can be compared with
the  curve 1 of Fig.7, which describes the same phase transition border
calculated in the one--loop approximation.
It is easy to see that the accuracy of the one--loop
approximation is not excellent and can commit errors of order 30\%.

According to the phase diagram drawn in Fig.7, the confinement phase
begins at $g^2 = g^2_{max}$ and exists under the phase transition border line
in the region $g^2 \le g^2_{max}$, where $e^2$ is large:
$e^2\ge (2\pi/g_{max})^2$, due to the Dirac relation (see
below).
Therefore, we have:
$$
g^2_{crit} = g^2_{max1}\approx 18.61
\quad -\quad
{\mbox{in the one--loop approximation}},\quad
$$
\begin{equation}
   g^2_{crit} = g^2_{max2}\approx
  15.11 \quad -\quad {\mbox{in the two--loop approximation}}.  \lb{52y}
\end{equation}
Comparing these results, we obtain the accuracy of
deviation between them of order 20\%.

The results (\ref{52y}) give:
\begin{equation}
   \tilde \alpha_{crit} = \frac {g^2_{crit}}{4\pi}\approx 1.48,
\quad -\quad{\mbox{in the one--loop approximation}},    \lb{53ay}
\end{equation}
\begin{equation}
   \tilde \alpha_{crit} = \frac {g^2_{crit}}{4\pi}\approx 1.20,
  \quad -\quad {\mbox{in the two--loop approximation}}.
                                                  \lb{53by}
\end{equation}
Using the Dirac relation for elementary charges:
\begin{equation}
   eg = 2\pi, \quad{\mbox{or}}\quad \alpha \tilde \alpha = \frac{1}{4},
                                             \lb{54y}
\end{equation}
we get the following values for the critical electric fine
structure constant:
\begin{equation}
        \alpha_{crit} = \frac{1}{4{\tilde \alpha}_{crit}}\approx 0.17
\quad -\quad{\mbox{in the one--loop approximation}},     \lb{55ay}
\end{equation}
\begin{equation}
        \alpha_{crit} = \frac{1}{4{\tilde \alpha}_{crit}}\approx 0.208
  \quad -\quad {\mbox{in the two--loop approximation}}.
                                               \lb{55by}
\end{equation}
The last result coincides with the lattice values (\ref{47}) obtained for the
compact QED by Monte Carlo method \ct{10s}.

Writing Eq.(\ref{24y}) with $\beta_g$ function given by Eqs.(\ref{33y}),
(\ref{34y}), and (\ref{37y}), we have the following RGE for the monopole
charge in the two--loop approximation:
\begin{equation}
  \frac{dg^2_{run}}{dt}\approx \frac{g^4_{run}}{48\pi^2} +
    \frac{g^6_{run}}{(16\pi^2)^2},        \lb{56ay}
\end{equation}
or
\begin{equation}
   \frac{d\log{\tilde \alpha}}{dt}\approx \frac{\tilde \alpha}{12\pi}
            (1 + 3\frac{\tilde \alpha}{4\pi}).    \lb{56by}
\end{equation}
The values (\ref{52y})  for $g^2_{crit} = g^2_{{max}1,2}$ indicate
that the contribution of two loops described by the second term
of Eq.(\ref{56ay}), or Eq.(\ref{56by}), is about 0.3, confirming the
validity of perturbation theory.

In general, we are able to estimate the validity of two--loop approximation
for all $\beta$--functions and $\gamma$, calculating the corresponding
ratios of two--loop contributions to one--loop contributions
at the maxima of curves 1 and 2:
\begin{equation}
\begin{array}{|l|l|}
\hline %
&\\[-0.2cm]%
\lambda_{crit} = \lambda_{run}^{max1} \approx{-13.16}&\lambda_{crit} =
\lambda_{run}^{max2}\approx{-7.13}\\[0.5cm]
g^2_{crit} = g^2_{max1}\approx{18.61}& g^2_{crit} = g^2_{max2}
\approx{15.11}\\[0.5cm]
\frac{\ds\gamma^{(2)}}{\ds\gamma^{(1)}}\approx{-0.0080}&\frac{\ds
\gamma^{(2)}}{\ds\gamma^{(1)}}\approx{-0.0065}\\[0.5cm]
\frac{\ds\beta_{\mu^2}^{(2)}}{\ds\beta_{\mu^2}^{(1)}}\approx{-0.0826}
&\frac{\ds\beta_{\mu^2}^{(2)}}{\ds\beta_{\mu^2}^{(1)}}
\approx{-0.0637}\\[0.8cm]
\frac{\ds\beta_{\lambda}^{(2)}}{\ds\beta_{\lambda}^{(1)}}\approx{0.1564}
&\frac{\ds\beta_{\lambda}^{(2)}}
{\ds\beta_{\lambda}^{(1)}}\approx{0.0412}\\[0.8cm]
\frac{\ds\beta_g^{(2)}}{\ds\beta_g^{(1)}}\approx{0.3536}&\frac{\ds
\beta_g^{(2)}}{\ds\beta_g^{(1)}}\approx{0.2871}\\[0.5cm]
\hline
\end{array}
                                         \lb{57y}
\end{equation}
Here we see that all ratios are sufficiently small, i.e. all
two--loop contributions are small in comparison with one--loop contributions,
confirming the validity of perturbation theory in the 2--loop
approximation, considered in this model. The accuracy of deviation is worse
($\sim 30\%$) for $\beta_g$--function. But it is necessary to emphasize
that calculating the border curves 1 and 2 of Fig.7, we have not used
RGE (\ref{37y}) for monopole charge: $\beta_g$--function is absent in
Eq.(\ref{49y}). Therefore, the calculation of $g^2_{crit}$ according to
Eq.(\ref{51y}) does not depend on the approximation of $\beta_g$ function.
The above--mentioned $\beta_g$--function appears only in the second order
derivative of $V_{eff}$ which is related with the monopole mass $m$
(see the next Section).

Eqs.(\ref{47}) and (\ref{55by}) give the result (\ref{50b}):
\begin{equation}
 \alpha_{crit}^{-1}\approx 5,      \lb{56y}
\end{equation}
which is important for the phase transition at the Planck scale
predicted by the Multiple Point Model (MPM) (see below).


\section{Approximate Universality of the Critical Coupling Constants}

The review of all existing results for $\alpha_{crit}$
and ${\tilde \alpha}_{crit}$ gives:

1)
\begin{equation}
    \alpha_{crit}^{lat}\approx{0.20 \pm 0.015},\quad
    {\tilde \alpha}_{crit}^{lat}\approx{1.25 \pm 0.10}     \lb{1u}
\end{equation}
-- in the Compact QED with the Wilson lattice action \ct{10s};

2)
\begin{equation}
    \alpha_{crit}^{lat}\approx{0.204} \quad
    {\tilde \alpha}_{crit}^{lat}\approx{1.25}    \lb{2u}
\end{equation}
-- in the model with the Wilson loop action \ct{19p};

3)
\begin{equation}
   \alpha_{crit} \approx 0.1836,\quad \tilde \alpha_{crit} \approx 1.36
                                                  \lb{3u}
\end{equation}
-- in the Compact QED with the Villain lattice action \ct{29y};

4)
\begin{equation}
     \alpha_{crit} = \alpha_{(A)}\approx{0.208},\quad
     {\tilde \alpha}_{crit} = {\tilde \alpha}_{(A)}\approx 1.20
                                     \lb{4u}
\end{equation}
-- in the HMM (\ct{22p},\ct{23p} and the present paper).

\vspace{0.1cm}
It is necessary to emphasize that the functions $\alpha (\beta)$
in Fig.5(a), describing the behavior of the effective electric fine structure
constant in the vicinity of the phase transition point, are different
for the Wilson and Villain lattice actions in the U(1) lattice gauge theory,
but the critical values of $\alpha$ coincide for both theories \ct{10s}.

Hereby we see an additional arguments for our previously hoped (in Refs.
\ct{17p} and \ct{19p}) "approximate universality" of the first order
phase transition couplings: the fine structure constant is
at the critical point approximately the same one independent of various
parameters of the different (lattice, etc.) regularization.

The most significant conclusion for MPM, predicting the values of gauge
couplings being so as to arrange just the multiple critical point where
all phases existing in the theory meet, is possibly that our calculations
suggest the validity of an approximate universality of the critical couplings,
in spite of the fact that we are concerned with the first order phase
transitions. We have shown that one can crudely
calculate the phase transition couplings without using any specific lattice,
rather only approximating the lattice artifact monopoles as fundamental
(pointlike) particles condensing. The details of the lattice -- hypercubic
or random, with multiplaquette terms or without them, etc., -- also the
details of the regularization -- lattice or Wilson loops, lattice or HMM --
do not matter for the value of the phase transition coupling so much.
Critical couplings depend only on groups with any regularization.
Such an approximate universality is, of course, absolutely needed if
there is any sense in relating lattice phase transition couplings to the
experimental couplings found in Nature. Otherwise, such a comparison would
only make sense if we could guess the true lattice in the right model,
what sounds too ambitious.

\section{Triple Point}

In this section we demonstrate the existence of the triple point on the
phase diagram of HMM.

Considering the second derivative of the effective potential:
\begin{equation}
                V''_{eff}(\phi^2)
    \equiv \frac{\partial^2 V_{eff}}{\partial {(\phi^2)}^2},
                                                 \lb{58y}
\end{equation}
we can calculate it for the RG improved effective potential (\ref{20y}):
$$
{V''}_{eff}(\phi^2) = \frac {{V'}_{eff}(\phi^2)}{\phi^2} + \biggl( - \frac 12
 \mu^2_{run}
+ \frac 12 \frac{d^2\mu^2_{run}}{dt^2} + 2\frac{d\mu^2_{run}}{dt}
 \frac{d\log G}{dt}
$$
$$
 +  \mu^2_{run}\frac{d^2\log G}{dt^2} +
   2\mu^2_{run}{(\frac{d\log G}{dt})}^2\biggl)\frac {G^2}{\phi^2} + \biggl(
   \frac 12 \frac{d\lambda_{run}}{dt} + \frac 14 \frac {d^2\lambda_{run}}
  {dt^2} + 2\frac{d\lambda_{run}}{dt}\frac{d\log G}{dt}
$$
\begin{equation}
 + 2\lambda_{run}\frac{d\log G}{dt} + \lambda_{run}\frac{d^2\log G}{dt^2} +
     4\lambda_{run}{(\frac{d\log G}{dt})}^2\biggl) G^4(t).
                                                    \lb{59y}
\end{equation}
Let us consider now the case when this second derivative
changes its sign giving a maximum of $V_{eff}$ instead of the minimum
at $\phi^2 = \phi_0^2$. Such a possibility is shown in Fig.6 by
the dashed curve "2". Now the two additional minima at $\phi^2 = \phi_1^2$
and $\phi^2 = \phi_2^2$ appear in our theory. They correspond to the two
different confinement phases for the confinement of electrically  charged
particles if they exist in the system. When these two minima are degenerate,
we have the following requirements:
\begin{equation}
       V_{eff}(\phi_1^2) = V_{eff}(\phi_2^2) < 0\quad {\mbox{and}}\quad
        {V'}_{eff}(\phi_1^2) = {V'}_{eff}(\phi_2^2) = 0,   \lb{61y}
\end{equation}
which describe the border between the confinement phases "conf.1" and "conf.2"
presented in Fig.8. This border is given as a curve "3" at the phase
diagram $(\lambda_{run}; g^4_{run})$ drawn in Fig.8. The curve "3"
meets the curve "1" at the triple point A.
According to the illustration shown in Fig.6, it is obvious
that this triple point A is given by the following requirements:
\begin{equation}
    V_{eff}(\phi_0^2) = V'_{eff}(\phi_0^2) = V''_{eff}(\phi_0^2) = 0.
                                         \lb{62y}
\end{equation}
In contrast to the requirements:
\begin{equation}
       V_{eff}(\phi_0^2) = V'_{eff}(\phi_0^2) = 0,    \lb{63y}
\end{equation}
giving the curve "1", let us consider now the joint solution of the following
equations:
\begin{equation}
         V_{eff}(\phi_0^2) = V''_{eff}(\phi_0^2) = 0 .    \lb{64y}
\end{equation}
For simplicity, we have considered the one--loop approximation.
Using Eqs.(\ref{59y}), (\ref{42y}) and (\ref{31y})-(\ref{33y}),
it is easy to obtain the solution of Eq.(\ref{64y}) in the one--loop
approximation:
\begin{equation}
            {\cal F}(\lambda_{run}, g^2_{run}) = 0,
                                                            \lb{65y}
\end{equation}
where
$$
{\cal F}(\lambda_{run}, g^2_{run}) = 5g_{run}^6 +
  24\pi^2g_{run}^4 + 12\lambda_{run}g_{run}^4 - 9\lambda_{run}^2g_{run}^2
$$
\begin{equation}
  + 36\lambda_{run}^3 + 80\pi^2\lambda_{run}^2 + 64\pi^4\lambda_{run}.
                                                \lb{66y}
\end{equation}
The dashed curve "2" of Fig.8 represents the solution of
Eq.(\ref{65y}) which is equivalent to Eqs.(\ref{64y}). The curve "2"
is going very close to the maximum of the curve "1".
Assuming that the position of the triple point A
coincides with this maximum,
let us consider the border between the phase "conf.1", having the first
minimum at nonzero $\phi_1$  with $V_{eff}^{min}(\phi_1^2) = c_1 < 0$,
and the phase "conf.2 ", which reveals two minima with the second minimum
being the deeper one and having $V_{eff}^{min}(\phi_2^2)=c_2 < 0$.
This border (described by the curve "3" of Fig.8) was calculated in the
vicinity of the triple point A by means of Eqs.(\ref{61y})
with $\phi_1$ and $\phi_2$ represented as $ \phi_{1,2} = \phi_0 \pm \epsilon$
with $\epsilon << \phi_0$. The result of such calculations gives the
following expression for the curve "3":
\begin{equation}
  g^4_{PT} = \frac {5}{2} ( 5\lambda_{run} + 8 \pi^2) \lambda_{run} + 8\pi^4.
                                                 \lb{67y}
\end{equation}
The curve "3" meets the curve "1" at the triple point A.

The piece of the curve "1" to the left of the point A describes the border
between the Coulomb--like phase and the phase "conf.1". In the vicinity
of the triple point A the second derivative $V_{eff}''(\phi_0^2)$ changes
its sign leading to the existence of the maximum at $\phi^2=\phi_0^2$,
in correspondence with the dashed curve "2" of Fig.6.  By this reason,
the curve "1" of Fig.8 does not already describe a phase transition border
up to the next point B when the curve "2" again intersects the curve "1"
at $\lambda_{(B)}\approx - 12.24$. This intersection (again giving
$V''_{eff}(\phi_0^2) > 0$) occurs surprisingly quickly.

The right piece of the curve "1" to the right of the point B
separates the Coulomb--like phase and the phase "conf.2". But
between the points A and B the phase transition border is going
slightly above the curve "1". This deviation is very small and cannot be
distinguished on Fig.8.

It is necessary to note that only $V''_{eff}(\phi^2)$ contains the derivative
$dg^2_{run}/dt$. The joint solution of equations (\ref{62y})
leads to the joint solution of Eqs.(\ref{50y}) and (\ref{65y}).
This solution was obtained numerically and gave the following triple point
values of $\lambda_{run}$ and $g^2_{run}$:
\begin{equation}
    \lambda_{(A)}\approx{ - 13.4073},\quad
              g^2_{(A)}\approx{18.6070}.            \lb{68y}
\end{equation}
The solution (\ref{68y}) demonstrates that the triple point A
exists in the very neighborhood of the maximum of the curve (\ref{50y}).
The position of this maximum is given by the following analytical
expressions, together with their approximate values:
\begin{equation}
     \lambda_{(A)}\approx - \frac{4\pi^2}3\approx -13.2,
                                                       \lb{69y}
\end{equation}
\begin{equation}
     g^2_{(A)} = g^2_{crit}|_{\mbox{for}\;\lambda_{run}=\lambda_{(A)}}
                 \approx \frac{4\sqrt{2}}3{\pi^2}\approx 18.6.
                                                        \lb{70y}
\end{equation}
Finally, we can conclude that the phase diagram shown in Fig.8 gives
such a description: there exist three phases in the dual sector of the Higgs
scalar electrodynamics -- the Coulomb--like phase and confinement
phases "conf.1" and "conf.2".

The border "1", which is described by the curve (\ref{50y}), separates
the Coulomb--like phase (with $V_{eff} \ge 0$) and confinement phases
(with $V_{eff}^{min}(\phi_0^2) < 0$).
The curve "1" corresponds to the joint solution of the equations
$V_{eff}(\phi_0^2)=V'_{eff}(\phi_0^2)=0$.

The dashed curve "2" represents the solution of the equations
$V_{eff}(\phi_0^2)=V''_{eff}(\phi_0^2)=0$.

The phase border "3" of Fig.8 separates two confinement phases.
The following requirements take place for this border:
$$
          V_{eff}(\phi_{1,2}^2) < 0,\qquad
         V_{eff}(\phi_1^2) = V_{eff}(\phi_2^2),\qquad
         V'_{eff}(\phi_1^2) = V'_{eff}(\phi_2^2) = 0,
$$
\begin{equation}
         V''_{eff}(\phi_1^2) > 0,\qquad V''_{eff}(\phi_2^2) > 0.
                                               \lb{71y}
\end{equation}
The triple point A is a boundary point of all three phase transitions
shown in the phase diagram of Fig.8.
For $g^2 < g^2_{({A})}$ the field system, described by our model, exists
in the confinement phase, where all electric charges  have to be confined.

Taking into account that monopole mass $m$ is given by the
following expression:
\begin{equation}
  \frac{d^2V_{eff}}{d\phi^2}|_{\phi=\phi_0} = m^2,                \lb{72y}
\end{equation}
we see that monopoles acquire zero mass in the vicinity of the triple point A:
\begin{equation}
  V''_{eff}(\phi_{0A}^2) =
\frac 1{4\phi_{0A}^2}\frac {d^2V_{eff}}{d\phi^2}|_{\phi=\phi_{0A}}
 = \frac {m^2_{(A)}}{4\phi_{0A}^2} = 0.
                               \lb{74y}
\end{equation}
This result is in agreement with the result of compact QED \ct{29y}:
$m^2\to 0$ in the vicinity of the critical point.

\section {"ANO--Strings", or the Vortex Description of the
Confinement Phases}

As it was shown in the previous subsection, two regions between the curves
"1", "3" and "3", "1", given by the phase diagram of Fig.8, correspond to the
existence of the two confinement phases, different in the sense
that the phase "conf.1" is produced by the second minimum, but the phase
"conf.2" corresponds to the third minimum of the effective potential.
It is obvious that in our case both phases have nonzero monopole condensate
in the minima of the effective potential, when
$V_{eff}^{min}(\phi_{1,2}\neq 0) < 0$. By this reason, the
Abrikosov--Nielsen--Olesen (ANO) electric vortices
(see Refs.\ct{30y},\ct{31y}) may exist in these both phases, which are
equivalent in the sense of the "string" formation.  If electric
charges are present in a model (they are absent in HMM), then these charges
are placed at the ends of the vortices--"strings" and therefore are
confined. But only closed "strings" exist in the confinement phases of HMM.
The properties of the ANO--"strings" in the U(1) gauge theory were
investigated in Ref.\ct{21p}.

As in Refs.\ct{30y},\ct{31y}, in the London's limit ($\lambda \to \infty$)
the dual Abelian Higgs model, described by the Lagrangian (\ref{5y}), gives
the formation of monopole condensate with amplitude $\phi_0$, which repels
and suppresses the electromagnetic field $F_{\mu\nu}$ almost everywhere,
except the region around the vortex lines. In this limit, we have the
following London equation:
\begin{equation}
      {\mbox rot}\; {\vec j}^{(m)} = {\delta}^{-2}{\vec E},      \lb{112x}
\end{equation}
where ${\vec j}^{(m)}$ is the microscopic current of monopoles,
${\vec E}$ is the electric field strength and $\delta$ is the penetration
depth. It is clear that ${\delta}^{-1}$ is the photon mass $m_V$,
generated by the Higgs mechanism. The closed equation for $\vec E$
follows from the Maxwell equations and Eq.(\ref{112x}) just in the
London's limit.

In our case $\delta$ is defined by the following relation:
\begin{equation}
            {\delta}^{-2} \equiv m_V^2 = g^2 \phi_0^2.    \lb{113x}
\end{equation}
On the other hand, the field $\phi$ has its own correlation length
$\xi$, connected to the mass of the field $\phi$ ("the Higgs mass"):
\begin{equation}
        \xi = {m_S}^{-1}, \qquad m_S^2 = \lambda \phi_0^2.   \lb{114x}
\end{equation}
The London's limit for our "dual superconductor of the second kind"
corresponds to the following relations:
\begin{equation}
      \delta >> \xi,\qquad m_V << m_S,\qquad g << \lambda,    \lb{115x}
\end{equation}
and "the string tension" -- the vortex energy per unit length (see
Ref.\ct{30y}) -- for the minimal electric vortex flux $2\pi$, is:
\begin{equation}
       \sigma = \frac {2\pi}{g^2\delta^2}\ln \frac{\delta}{\xi}
              = 2\pi \phi_0^2 \ln \frac{m_S}{m_V}, \qquad {\mbox
        {where}}\qquad \delta/\xi = \frac {m_S}{m_V} >> 1.      \lb{116x}
\end{equation}
We see that in the London's limit ANO--theory implies the mass generation
of the photons, $m_V = 1/\delta$, which is much less than the Higgs
mass $m_S = 1/\xi$.

Let us wonder now in the question whether our "strings" are
thin or not. The vortex may be considered as thin, if the distance
between the electric charges sitting at its ends, i.e. the string length $L$,
is much larger than the penetration length $\delta$:
\begin{equation}
            L >> \delta >> \xi.         \lb{117x}
\end{equation}
It is obvious that only rotating "strings" can exist as stable states.
In the framework of classical calculations, it is not difficult to
obtain the mass $M$ and angular momentum $J$ of the rotating "string":
\begin{equation}
             J = \frac{1}{2\pi \sigma} M^2, \qquad
             M = \frac {\pi}{2} \sigma L.               \lb{118x}
\end{equation}
The following relation follows from Eqs.(\ref{118x}):
\begin{equation}
        L = 2 \sqrt{\frac{2J}{\pi \sigma}},             \lb{119x}
\end{equation}
or
\begin{equation}
       L = \frac{2g\delta}{\pi}\sqrt{\frac{J}{\ln \frac{m_S}{m_V}}}.
                             \lb{120x}
\end{equation}
For $J=1$ we have:
\begin{equation}
       \frac{L}{\delta} = \frac{2g}{\pi \sqrt{\ln \frac {m_S}{m_V}}},
                                                            \lb{121x}
\end{equation}
what means that for $ m_S >> m_V $ the length of this "string" is small
and does not obey the requirement (\ref{117x}). It is easy to see from
Eq.(\ref{120x}) that in the London's limit the "strings" are very thin
($L/\delta >> 1$) only for the enormously large angular momenta $J >> 1$.

The phase diagram of Fig.8 shows the existence of the confinement phase
for $\alpha \ge \alpha_{(A)}$. This means that the formation
of (closed) vortices begins at the triple point $\alpha = \alpha_{(A)}$:
for $\alpha > \alpha_{(A)}$ we have a nonzero $\phi_0$ giving rise to
existence of vortices.

In Section 3 we have shown that the lattice investigations lead to the
freezing of the electric fine structure constant at the value
$\alpha = \alpha_{max}$ and mentioned that the authors of Ref.\ct{14p}
predicted: $\alpha_{max}=\frac {\pi}{12} \approx 0.26$.

Let us estimate now the region of values of the magnetic
charge $g$ in the  confinement phase considered in this paper:
$$
               g_{min} \le g \le g_{max},
$$
$$
           g_{max} = g_{(A)}\approx{\sqrt {15.1}}\approx 3.9,
$$
\begin{equation}
           g_{min} = \sqrt {\frac {\pi}{\alpha_{max}}}\approx 3.5.
                                             \lb{122x}
\end{equation}
Then for $m_S = 10m_V$, say, we have from Eq.(\ref{121x}) the following
estimation of the "string" length for $J=1$:
\begin{equation}
            1.5 \stackrel{<}{\sim}\frac{L}{\delta}\stackrel{<}{\sim} 1.8.
                              \lb{123x}
\end{equation}
We see that in the $U(1)$ gauge theory the low-lying states of "strings" correspond to the short
and thick vortices.

In general, the way of receiving of the Nambu--Goto strings
from the dual Abelian Higgs model of scalar monopoles was
demonstrated in Ref.\cite{32y}.

\section{Phase Transition Couplings in the Regularized SU(N) Gauge Theories}

It was shown in a number of investigations (see for example \ct{11s},\ct{12s}
and references there), that the confinement in the SU(N) lattice gauge
theories effectively comes to the same U(1) formalism. The reason is the
Abelian dominance in their monopole vacuum: monopoles of the Yang--Mills
theory are the solutions of the U(1)--subgroups, arbitrary embedded into
the SU(N) group. After a partial gauge fixing (Abelian projection by
't Hooft \ct{24p}) SU(N) gauge theory is reduced to an Abelian
$U(1)^{N-1}$ theory with $N-1$ different types of Abelian monopoles.
Choosing the Abelian gauge for dual gluons, it is possible to describe
the confinement in the lattice SU(N) gauge theories by the analogous
dual Abelian Higgs model of scalar monopoles.

\subsection{The "abelization" of monopole vacuum in the non-Abelian theories}

A lattice imitates the non--perturbative vacuum of zero temperature
SU(2) and SU(3) gluodynamics as a condensate of monopoles which emerge
as leading non--perturbative fluctuations of non--Abelian
SU(N) gauge theories in the gauge of Abelian projections
by G.'t Hooft \ct{24p}.
It is possible to find such a gauge, in which monopole degrees of
freedom, hidden in the given field configuration, become explicit.

Let us consider the SU(N) gluodynamics. For any composite operator
$\bf X\in$ {the adjoint representation of SU(N) group}
($\bf X$ may be ${(F_{\mu\nu})}_{ij}$, where $i,j=1,2,...N$)
we can find such a gauge:
\begin{equation}
{\bf    X \to X' = VXV^{-1},           }                \lb{1h}
\end{equation}
where the unitary matrix $\bf V$ transforms $\bf X$ to diagonal $\bf X'$:
\begin{equation}
{\bf    X \to X' = VXV^{-1} = {\mbox{diag}}(
\lambda_1, \lambda_2,..., \lambda_N).           }          \lb{2h}
\end{equation}
We can choose the ordering of $\lambda_i$:
\begin{equation}
            \lambda_1 \le \lambda_2 \le...\le \lambda_N.         \lb{3h}
\end{equation}
The matrix $\bf X'$ belongs to the Cartain, or Maximal Abelian subgroup
of the SU(N) group:
\begin{equation}
             U(1)^{N-1}\in SU(N).                          \lb{4h}
\end{equation}
Let us consider the field $A_{\mu}$ in the diagonal gauge:
\begin{equation}
          {\bar A}_{\mu} =
V ( A_{\mu} + \frac{i}{g}\partial_{\mu})V^{-1}.
                                      \lb{5h}
\end{equation}
This field transforms according to the subgroup $U(1)^{N-1}$: its diagonal elements
$$
{(a_{\mu})}_i
\equiv {({\bar A}_{\mu})}_{ii}
$$
transform as Abelian gauge fields (photons):
\begin{equation}
{(a_{\mu})}_i \to {(a'_{\mu})}_i = {(a_{\mu})}_i +
              \frac{1}{g}\partial_{\mu} \alpha_i,
                                        \lb{6h}
\end{equation}
but its non--diagonal elements
$$
{(c_{\mu})}_{ij}
\equiv {({\bar A}_{\mu})}_{ij}\quad {\mbox{with}}\quad i\neq j
$$
transform as charged fields:
\begin{equation}
{(c'_{\mu})}_{ij} = \exp[i(\alpha_i - \alpha_j)]{(c_{\mu})}_{ij},
                                        \lb{7h}
\end{equation}
where $i,j = 1,...,N$.

According to G.'t Hooft~\ct{24p}, if some $\lambda_i$ coincide, then
singularities having properties of monopoles appear in the "Abelian"
part of the non--Abelian gauge fields. Indeed, let us consider the
strength tensor of the "Abelian gluons":
$$
{(f_{\mu\nu})}_i = \partial_{\mu}{(a_{\nu})}_i - \partial_{\nu}{(a_{\mu})}_i
$$
\begin{equation}
 = V F_{\mu\nu}V^{-1} + ig[
V ( A_{\mu} + \frac{i}{g}\partial_{\mu})V^{-1},
V ( A_{\nu} + \frac{i}{g}\partial_{\nu})V^{-1}].
                                      \lb{8h}
\end{equation}
The monopole current is:
\begin{equation}
   {(K_{\mu})}_i = \frac{1}{8\pi}\epsilon_{\mu\nu\rho\sigma}
\partial_{\nu}{(f_{\rho\sigma})}_i,
                                        \lb{9h}
\end{equation}
and it is conserved:
\begin{equation}
    \partial_{\mu}{(K_{\mu})}_i = 0.
                                 \lb{9ah}
\end{equation}
The initial $F_{\mu\nu}$ had not singularities. Therefore, all
singularities can come from the commutator, which is contained in
Eq.(\ref{8h}).

The magnetic charge $m_i(\Omega)$ in 3d--volume $\Omega $ is:
\begin{equation}
  m_i(\Omega)= \int_{\Omega} d^3 \sigma_{\mu}{(K_{\mu})}_i =
   \frac{1}{8\pi}\int_{\partial{\Omega}} d^2 \sigma_{\mu\nu}
    {(f_{\mu\nu})}_i.
                                    \lb{10h}
\end{equation}
If $\lambda_1\,=\,\lambda_2$ (coincide) at the point $x^{(1)}$
in 3d--volume $\Omega$, then we have a singularity on the curve
in 4d--space, which is a world--line of the magnetic monopole,
and $x = x^{(1)}$ is a singular point of gauge transformed
fields ${\bar A}_{\mu}\,$ and $\,{(a_{\mu})}_i$.

As it was shown by 't Hooft:
\begin{equation}
{(f_{\mu\nu})}_i\sim O(|x - x^{(1)}|^{-2})              \lb{11h}
\end{equation}
only in the vicinity of $x^{(1)}$, where it behaves as a magnetic field
of the point--like monopole.

Finally, we have the following conclusions:

1) The initial potentials $A_{\mu}\,$ and strength tensor
$F_{\mu\nu}$ had not singularities.

2) At large distances ${(f_{\mu\nu})}_i$ doesn't have a behaviour
$$
{(f_{\mu\nu})}_i \sim O(|x - x^{(1)}|^{-2})
$$
and we have monopoles only near $x = x^{(1)}$.

3) Fields ${\bar A}_{\mu}\,$ and $\,{(a_{\mu})}_i$ are not
classical solutions: they are a result of the quantum fluctuations.

4) Any distribution of fields in the vacuum can undergo the
Abelian projection.

We have seen that in the SU(N) gauge theories quantum fluctuations
(non--perturbative effects) reveal an Abelian vacuum monopoles
and suppress the non--diagonal components of the strength tensor
${(F_{\mu\nu})}_{ij}$.
As it is shown below, this phenomenon gives very important
consequences for the Planck scale physics.

Using the idea of "abelization" of monopole vacuum in the
SU(N) lattice gauge theories, we have developed in Ref.\ct{22p} a method of
theoretical estimation of the SU(N) critical couplings.

\subsection{Monopoles strength group dependence}

Lattice non--Abelian gauge theories also have lattice artifact monopoles.
We suppose that only those lattice artifact monopoles are important for the
phase transition calculations which have the smallest monopole charges.
Let us consider the lattice gauge theory with the gauge group $SU(N)/Z_N$
as our main example.
That is to say, we consider the adjoint representation action and do not
distinguish link variables forming the same one multiplied by any element of
the center of the group. The group $ SU(N)/Z_N $ is not simply connected
and has the first homotopic group $\Pi_1(SU(N)/Z_N)$ equal to $Z_N$.
The lattice artifact monopole with the smallest magnetic charge may be
described as a three--cube (or rather a chain of three--cubes describing the
time track) from which radiates magnetic field corresponding to the $U(1)$
subgroup of gauge group $SU(N)/Z_N$ with the shortest length insight of
this group, but still homotopically non-trivial. In fact, this $U(1)$
subgroup is obtained by the exponentiating generator:
\begin{equation}
\frac{``\lambda_8``}{2} = \frac {1}{\sqrt{2N(N-1)}}
\left(\begin{array}{*{4}{c}}
N-1 & 0 & \cdots & 0\\
0 & -1 & \cdots & 0 \\
\vdots & \vdots &\ddots & \vdots\\
0 & 0 &  \cdots & -1 \\
\end{array}
\right).
\lb{1z}
\end{equation}
This specific form is one gauge choice; any similarity transformation
of this generator would describe physically the same monopole. If one
has somehow already chosen the gauge monopoles with different but similarity
transformation related generators, they would be physically different.
Thus, after gauge choice, there are monopoles corresponding to different
directions of the Lie algebra generators in the form ${\cal
U}\frac{``\lambda_8``}{2}\,{\cal U}^{+}$.

Now, when we  want to apply the effective potential calculation as a
technique for the getting phase diagram information for the condensation
of the lattice artifact monopoles in the non-abelian lattice gauge theory,
we have to correct the abelian case calculation for the fact that
after gauge choice we have a lot of different monopoles.
If a couple of monopoles happens to have their generators just in the same
directions in the Lie algebra, they will interact with each other as
Abelian monopoles (in first approximation). In general, the interaction
of two monopoles by exchange of a photon will be modified by the
following factor:
\begin{equation}
  \frac{ Tr ({\cal U}_1\frac{``\lambda_8``}{2} {\cal U}_1^{+}\,
{\cal U}_2\frac{``\lambda_8``}{2} {\cal U}_2^{+})}
{Tr{(\frac{``\lambda_8``}2)}^2}.          \lb{2z}
\end{equation}
We shall assume that we can correct these values of monopole orientations in
the Lie algebra in a statistical way. That is to say, we want to determine
an effective coupling constant ${\tilde g}_{eff}$ describing the
monopole charge as if there is only one Lie algebra orientationwise type
of monopole. It should be estimated statistically in terms of the monopolic
charge ${\tilde g}_{genuine}$ valid to describe the interaction between
monopoles with generators oriented along the same $U(1)$ subgroup.
A very crude intuitive estimate of the relation between these two
monopole charge concepts ${\tilde g}_{genuine}$ and ${\tilde g}_{eff}$
consists in playing that the generators are randomly oriented in the
whole $N^2 - 1$ dimensional Lie algebra. When even the sign of the Lie
algebra generator associated with the monopole is random -- as we assumed
in this crude argument -- the interaction between two monopoles with
just one photon exchanged averages out to zero. Therefore, we can get
a non-zero result only in the case of exchange by two photons or more.
That is, however, good enough for our effective potential calculation
since only ${\tilde g}^4$ (but not the second power) occurs in the
Coleman -- Weinberg effective potential in the one--loop approximation
(see \ct{20s},\ct{21s}). Taking into account this fact that we can
average imagining monopoles with generators along a basis vector in the
Lie algebra, the chance of interaction by double photon exchange between
two different monopoles is just $\frac{1}{N^2 - 1}$, because there are
$N^2 - 1$ basis vectors in the basis of the Lie algebra. Thus, this crude
approximation gives:
\begin{equation}
    {\tilde g}^4_{eff} = \frac{1}{N^2 - 1}{\tilde g}^4_{genuine}.
                                     \lb{3z}
\end{equation}

Note that considering the two photons exchange which is forced by our
statistical description, we must concern the forth power
of the monopole charge $\tilde g$.

The relation (\ref{3z}) was not derived correctly, but its
validity can be confirmed if we use a more correct statistical argument.
The problem with our crude estimate is that the generators making monopole
charge to be minimal must go along the shortest type of U(1) subgroups with
non-trivial homotopy.

\subsubsection*{Correct averaging}

The $\lambda_8$--like generators
${\cal U}\frac{``\lambda_8``}{2}\,{\cal U}^+$  maybe written as
\begin{equation}
{\cal U} \frac{``\lambda_8``}{2} \,{\cal U}^{+}
= - \sqrt{\frac {1}{2N(N-1)}} {\bf 1\!\!\! 1} + \sqrt{\frac{N}{2(N-1)}}
\,{\cal P},
\lb{4z}
\end{equation}
where $\cal P$ is a projection metrics into one--dimensional state in
the \un{N} representation. It is easy to see that averaging according to
the Haar measure distribution of ${\cal U}$, we get the average of
$\cal P$ projection on ``quark'' states with a distribution corresponding
to the rotationally invariant one on the unit sphere in the N--dimensional
\un{N}--Hilbert space.

If we denote the Hilbert vector describing the state
on which $\cal P$ shall project as
\begin{equation}
{\cal P} =\left(
\begin{array}{c}
\psi_1\\
\psi_2\\
\vdots\\
\psi_N
\end{array}
\right),
                                       \lb{5z}
\end{equation}
then the probability distribution on the unit sphere becomes:
\begin{equation}
P (\left(
\begin{array}{c}
\psi_1\\
\psi_2\\
\vdots\\
\psi_N
\end{array}
\right)) {\prod}_{i=1}^N d{\psi}_i\propto \delta({\sum}_{i=1}^N {|{\psi}_i|}^2
- 1) \prod_{i=1}^N d({|\psi_i|}^2).               \lb{6z}
\end{equation}
Since, of course, we must have ${|{\psi}_i|}^2 \ge 0$ for all $i=1,2,...,N$,
the $\delta$--function is easily seen to select a flat distribution
on a (N - 1)--dimensional equilateral simplex. The average of the two
photon exchange interaction given by the
correction factor (\ref{2z}) squared (numerically):
\begin{equation}
\frac{Tr ({\cal U}_1 \frac{``\lambda_8``}{2} {\cal U}_1^{+}{\cal U}_2
{\frac{``\lambda_8``}{2}}{\cal U}_2^{+})^2}{Tr({(\frac{``\lambda_8``}{2})}^2)
^2}
                                 \lb{7z}
\end{equation}
can obviously be replaced by the expression where we take as random only
one of the ``random``  $\lambda_8$--like generators, while the other one is
just taken as $\frac{``\lambda_8``}{2}$, i.e. we can take say
${\cal U}_2 = {\bf 1\!\!\!1}$ without changing the average.

Considering the two photon exchange diagram, we can write
the correction factor (obtained by the averaging)
for the fourth power of magnetic charge:
\begin{equation}
\frac{{\tilde g}^4_{eff}}{{\tilde g}^4_{{\rm genuine}}}
= {\rm average}\{\frac{Tr{(\frac{``\lambda_8 ``}{2} {{\cal U}_{1}}
{\frac{``\lambda_8``}{2}}{{\cal U}_{1}}^{+})}^2}
{Tr{({(\frac{``\lambda_8``}{2})}^2)}^2}\}.       \lb{8z}
\end{equation}
Substituting the expression (\ref{4z}) in Eq.(\ref{8z}), we have:
\begin{equation}
\frac{{\tilde g}^4_{eff}}{{\tilde g}^4_{{\rm genuine}}}
= {\rm average}\{\frac{ Tr{(\frac{``\lambda_8``}{2} (-\sqrt{\frac{1}
{2N(N-1)}}{\bf 1\!\!\!1}
 + \sqrt{\frac {N}{2(N-1)}}{\cal P}))}^2}
{Tr{({(\frac{``\lambda_8``}{2})}^2)}^2}\}.
                                            \lb{8z}
\end{equation}
Since $\frac{``\lambda_8``}{2}$ is traceless, we obtain using the
projection (\ref{5z}):

\begin{equation}
 Tr(\frac{``\lambda_8``}{2} {\cal P}\sqrt{\frac{N}{2(N-1)}})
= - \frac{1}{2(N-1)} + \frac{N}{2(N-1)}{|\psi_1|}^2.
                                              \lb{9z}
\end{equation}
The value of the square ${|\psi_1|}^2$ over the simplex is proportional
to one of the heights in this simplex. It is obvious from the geometry of a
simplex that the distribution of ${|\psi_1|}^2$ is
\begin{equation}
     d{\mbox P} = (N-1){(1 - {|\psi_1|}^2)}^{(N-2)} d({|\psi_1|}^2),
                                          \lb{10z}
\end{equation}
where, of course, $0 \le {|{\psi}_1|}^2 \le 1$ only is allowed.
In Eq.(\ref{10z}) P is a probability. By definition:
\begin{equation}
     {\rm average}\{f({|{\psi}_1|}^2)\}
= (N-1)\int_0^1 f({|{\psi}_1|}^2)(1 - {|\psi_1|}^2)^{(N-2)}d({|\psi_1|}^2).
                                            \lb{11z}
\end{equation}
Then
\begin{eqnarray}
\frac{{\tilde g}^4_{eff}}{{\tilde g}^4_{{\rm genuine}}}
&=& \frac{N^2}{(N-1)} \int_0^1
{(\frac{1}{N} - {|\psi_1|}^2)}^2{(1 - {|\psi_1|}^2)}^{N-2} d({|\psi_1|}^2)\\
&=& \frac{N^2}{N-1}\int_0^1 {(1-y-\frac{1}{N})}^2\,y^{N-2}dy\\
&=& \frac{1}{N^2-1}                \lb{12z}
\end{eqnarray}
and we have confirmed our crude estimation (\ref{3z}).

\subsubsection*{Relative normalization of couplings}

Now we are interested in how ${\tilde g}^2_{genuine}$ is related to
$\alpha_N=g^2_N/{4\pi}$.

We would get the simple Dirac relation:
\begin{equation}
          g_{(1)}\cdot {\tilde g}_{\rm{genuine}} = 2\pi,
                                              \lb{13z}
\end{equation}
if $g_{(1)}\equiv g_{U(1)-{\rm subgroup}}$ is the coupling for the
U(1)--subgroup of SU(N) normalized in such a way that the charge quantum
$g_{(1)}$ corresponds to a covariant derivative
$\partial_{\mu} - g_{(1)}\,A_{\mu}^{U(1)}$.

Now we shall follow the convention -- usually used to define
$\alpha_N={g^2_N}/{4\pi}$ -- that the covariant derivative for the
{\un{N}}--plet representation is:
\begin{equation}
        D_{\mu} = \partial_{\mu} - g_N \frac{\lambda^a}{2}A_{\mu}^a
                                                     \lb{14z}
\end{equation}
with
\begin{equation}
 Tr(\frac{\lambda^a}{2}\frac{\lambda^b}{2}) = \frac{1}{2}\delta^{ab},
                                                   \lb{15z}
\end{equation}
and the kinetic term for the gauge field is
\begin{equation}
       L = - \frac{1}{4}F_{\mu\nu}^aF^{a\,\mu\nu},   \lb{16z}
\end{equation}
where
\begin{equation}
    F_{\mu\nu}^a = \partial_{\mu}A_{\nu}^a - \partial_{\nu}A_{\mu}^a
                    - g_N\,f^{abc}A_{\mu}^bA_{\nu}^c.
                                         \lb{17z}
\end{equation}
Especially if we want to choose a basis for our generalized
Gell--Mann matrices so that one basic vector is our
$\frac{``\lambda_8``}{2}$, then for $A_{\mu}^{``8``}$ we have the
covariant derivation
$\partial_{\mu} - g_N\,\frac{``\lambda_8``}{2}\,A_{\mu}^{``8``}$.
If this covariant derivative is written in terms of the U(1)--subgroup,
corresponding to monopoles with the Dirac relation (\ref{13z}),
then the covariant derivative has a form
$\partial_{\mu} - g_{(1)}\,A_{\mu}^8\cdot{\un{\un{M}}}$. Here
${\un{\un{M}}}$ has the property that $\exp(i2\pi{\un{\un{M}}})$
corresponds to the elements of the group $SU(N)/Z_N$ going all around and
back to the unit element. Of course, ${\un{\un{M}}}
=\frac{g_N}{g_{(1)}}\cdot\frac{``\lambda_8``}{2}$ and the ratio
$g_N/g_{(1)}$ must be such one that
$\exp(i2\pi\frac{g_N}{g_{(1)}}\frac{``\lambda_8``}{2})$
shall represent -- after first return -- the unit element of the
group $SU(N)/Z_N$. Now this unit element really means the coset consisting
of the center elements $\exp(i\frac{2\pi k}{N})\in SU(N),\,(k\in Z)$, and
the requirement of the normalization of $g_{(1)}$ ensuring the Dirac
relation (\ref{13z}) is:
\begin{equation}
    \exp(i2\pi\frac{g_N}{g_{(1)}}\frac{``\lambda_8``}{2})
                      = \exp(i\frac{2\pi}{N}){\bf 1\!\!\!1}.
                                           \lb{18z}
\end{equation}
This requirement is satisfied if the eigenvalues of
$\frac{g_{N}}{g_{(1)}}\frac{``\lambda_8``}{2}$ are
modulo 1 equal to $-\frac{1}{N}$, i.e. formally we might write:
\begin{equation}
   \frac{g_{N}}{g_{(1)}}\frac{``\lambda_8``}{2}
     = - \frac{1}{N}\quad (\rm{mod}\, 1).                  \lb{19z}
\end{equation}
According to (\ref{1z}), we have:
\begin{equation}
\frac{g_N}{g_{(1)}}\cdot \frac {1}{\sqrt{2N(N-1)}}
\left(\begin{array}{*{4}{c}}
N-1 & 0 & \cdots & 0\\
0 & -1 & \cdots & 0 \\
\vdots & \vdots &\ddots & \vdots\\
0 & 0 &  \cdots & -1 \\
\end{array}
\right)
= - \frac{1}{N}\quad (\rm{mod}\, 1),                  \lb{20z}
\end{equation}
what implies:
\begin{equation}
   \frac{g_N}{g_{(1)}} = \sqrt{\frac{2(N-1)}{N}},         \lb{21z}
\end{equation}
or
\begin{equation}
   \frac{g_N^2}{g_{(1)}^2} = \frac{2(N-1)}{N}.         \lb{22z}
\end{equation}

\subsection{The relation between U(1) and SU(N) critical couplings}

Collecting the relations (\ref{22z}), (\ref{13z}) and (\ref{3z}),
we get:
$$
   \alpha_N^{-1} = \frac{4\pi}{g_N^2}
= \frac{N}{2(N-1)}\cdot \frac
    {4\pi}{g^2_{(1)}}
= \frac{N}{2(N-1)}\cdot \frac{{\tilde g}^2_{genuine}}{\pi}
$$
\begin{equation}
= \frac{N}{2(N-1)}\sqrt{N^2-1}\cdot \frac{{\tilde g}^2_{eff}}{\pi}
= \frac{N}{2(N-1)}\sqrt{N^2-1}\cdot \frac{4\pi}{g_{U(1)}^2}
= \frac{N}{2}\sqrt{\frac{N+1}{N-1}}\cdot \alpha_{U(1)}^{-1},
                                                            \lb{23z}
\end{equation}
where
\begin{equation}
          g_{U(1)}{\tilde g}_{eff} = 2\pi                 \lb{24z}
\end{equation}
and $\alpha_{U(1)}=g_{U(1)}^2/{4\pi}$.

The meaning of this result is that provided that we have
${\tilde g}_{eff}$ the same for $SU(N)/Z_N$ and U(1) gauge theories
the couplings are related according to Eq.(\ref{23z}).

We have a use for this relation when we want to calculate the phase
transition couplings considering the scalar monopole field responsible for
the phase transition in the gauge groups $SU(N)/Z_N$. Having in mind
the "Abelian" dominance in the SU(N) monopole vacuum, we must think
that ${\tilde g}_{eff}^{crit}$ coincides with $g_{crit}$ of the
U(1) gauge theory. Of course, here we have an approximation taking
into account only monopoles interaction and ignoring the relatively
small selfinteractions of the Yang--Mills fields. In this approximation
we obtain the same phase transition (triple point, or critical)
${\tilde g}_{eff}$--coupling which is equal to $g_{crit}$ of U(1)
whatever the gauge group SU(N) might be. Thus we conclude that
for the various groups $U(1)$ and $SU(N)/{Z_N}$, according to Eq.(\ref{23z}),
we have the following relation between the phase transition couplings:
\begin{equation}
      \alpha_{N,crit}^{-1}
           = \frac{N}{2}\sqrt{\frac{N+1}{N-1}}
                          \alpha_{U(1),crit}^{-1}.
                                            \lb{25z}
\end{equation}
Using the relation (\ref{25z}), we obtain:
\begin{equation}
    \alpha_{U(1),crit}^{-1} : \alpha_{2,crit}^{-1} : \alpha_{3,crit}^{-1}
           = 1 : \sqrt{3} : 3/\sqrt{2} = 1 : 1.73 : 2.12.
                                                            \lb{26z}
\end{equation}
These relations will be used below for the explanation of predictions of MPM.

\section{Multiple Point Model and Critical Values
of the U(1) and SU(N) Fine Structure Constants}

Investigating the phase transition in the dual Higgs monopole model,
we have pursued two objects. From one side, we had an aim to
explain the lattice results. But we had also another aim.

According to MPM, at the Planck scale there exists a multiple critical point,
which is a boundary point of the phase transitions in U(1), SU(2) and SU(3)
sectors of the fundamental regularized gauge theory G. It is natural to assume
that the objects responsible for these transitions are the physically existing
Higgs scalar monopoles, which have to be introduced into theory as fundamental
fields.  Our calculations indicate that the corresponding critical couplings
coincide with the lattice ones, confirming the idea of Ref.\ct{17p}.

The results of the present paper are very encouraging for the
Anti--Grand Unification Theory (AGUT), which always is used
in conjunction with the Multiple Point Model (MPM).

\subsection{G-theory, or Anti--grand unification theory (AGUT)}

Most efforts to explain the Standard Model (SM) describing well all
experimental results known today are devoted to Grand Unification
Theories (GUTs). The supersymmetric extension of the SM consists of taking the
SM and adding the corresponding supersymmetric partners \ct{32a}.  The Minimal
Supersymmetric Standard Model (MSSM) shows \ct{33a} the possibility of the
existence of the grand unification point at
\bc
$\mu_{GUT}\sim 10^{16}$ GeV.
\ec
Unfortunately, at present time experiment does not indicate any manifestation
of the supersymmetry. In this connection, the Anti--Grand Unification
Theory (AGUT) was developed in Refs.\ct{2a}, \ct{2e}-\ct{2k},
\ct{16p}-\ct{23p} and \ct{35}-\ct{38} as a realistic alternative to SUSY GUTs.
According to this theory, supersymmetry does not come into the existence
up to the Planck energy scale (\ref{1}).
The Standard Model (SM) is based on the group SMG described by Eq.(\ref{2}).
AGUT suggests that at the scale $\mu_G\sim \mu_{Pl}=M_{Pl}$
there exists the more fundamental group $G$ containing $N_{gen}$
copies of the Standard Model Group SMG:
\begin{equation}
G = SMG_1\times SMG_2\times...\times SMG_{N_{gen}}\equiv (SMG)^{N_{gen}},
                                                  \lb{76y}
\end{equation}
where $N_{gen}$ designates the number of quark and lepton generations.

If $N_{gen}=3$ (as AGUT predicts~\cite{2k}), then the fundamental gauge group G is:
\begin{equation}
    G = (SMG)^3 = SMG_{1st\;gen.}\times SMG_{2nd\;gen.}\times SMG_{3rd\;gen.},
                                        \lb{77y}
\end{equation}
or the generalized one:
\begin{equation}
         G_f = (SMG)^3\times U(1)_f,           \lb{78y}
\end{equation}
which was suggested by the fitting of fermion masses of the SM
(see Refs.\ct{35}).

Recently a new generalization of AGUT was suggested in Refs.\ct{37}:
\begin{equation}
           G_{\mbox{ext}} = (SMG\times U(1)_{B-L})^3,    \lb{79y}
\end{equation}
which takes into account the see--saw mechanism with right-handed neutrinos,
also gives the reasonable fitting of the SM fermion masses and describes
all neutrino experiments known today.

By reasons considered in the last Section of this review, we prefer
not to use the terminology "Anti-grand unification theory, i.e. AGUT",
but call the theory with the group of symmetry $G$, or $G_f$,
or $G_{ext}$, given by Eqs.(\ref{76y})-(\ref{79y}), as "G--theory",
because, as it will be shown below, we have a possibility of the
Grand Unification near the Planck scale using just this theory.

The group $G_f$ contains the following gauge fields:
$3\times 8 = 24$ gluons, $3\times 3 = 9$ W-bosons and $3\times 1 + 1 = 4$
Abelian gauge bosons. The group $G_{ext}$ contains:
$3\times 8 = 24$ gluons, $3\times 3 = 9$ W-bosons and $3\times 1 +
3\times 1 = 6$ Abelian gauge bosons.

At first sight, this ${(SMG)}^3\times U(1)_f$ group with its 37 generators
seems to be just one among many possible SM gauge group extensions.
However, it is not such an arbitrary choice. There are at least
reasonable requirements (postulates) on the gauge group G (or $G_f$,
or $G_{ext}$) which have uniquely to specify this group.  It should obey the
following postulates (the first two are also valid for SU(5) GUT):

\vspace{0.1cm}

1. G or $G_f$ should only contain transformations, transforming the
known 45 Weyl fermions ( = 3 generations of 15 Weyl particles each)
-- counted as left handed, say -- into each other unitarily,
so that G (or $G_f$) must be a subgroup of U(45): $G\subseteq U(45)$.

\vspace{0.1cm}

2. No anomalies, neither gauge nor mixed. AGUT assumes that only
straightforward anomaly cancellation takes place and forbids the
Green-Schwarz type anomaly cancellation \ct{39}.

\vspace{0.1cm}

3. AGUT should NOT UNIFY the irreducible representations under the SM
gauge group, called here SMG (see Eq.(\ref{2})).

\vspace{0.1cm}

4. G is the maximal group satisfying the above-mentioned postulates.

\vspace{0.1cm}

There are five Higgs fields named $\phi_{WS}$, S, W, T, $\xi$
in AGUT extended by Froggatt and Nielsen \ct{35} with
the group of symmetry $G_f$ given by Eq.(\ref{78y}).
These fields break AGUT to the SM what means that their vacuum expectation
values (VEV) are active. The field $\phi_{WS}$ corresponds
to the Weinberg--Salam theory, $<S>=1$, so that we have only
three free parameters -- three VEVs $\;<W>, <T>$ and $<\xi>$ to fit
the experiment in the framework of this model.
The authors of Refs.\ct{35} used them with
aim to find the best fit to conventional experimental data
for all fermion masses and mixing angles in the SM (see Table I).

The result is encouraging. The fit is given by the ${\chi}^2$ function
(called here ${\tilde \chi}^2$). The lowest value of ${\tilde \chi}^2
(\approx 1.87)$ gives the following VEVs:
\begin{equation}
<S>=1;\quad\quad<W>=0.179;\quad\quad <T>=0.071;\quad\quad <\xi >=0.099.
                                                     \lb{80y}    
\end{equation}
The extended AGUT by Nielsen and Takanishi \ct{37}, having
the group of symmetry $G_{ext}$ (see Eq.(\ref{79y})),
was suggested with aim to explain the neutrino oscillations.
Introducing the right--handed neutrino in the model, the authors replaced the
assumption 1 and considered U(48) group instead of U(45), so that
$G_{ext}$ is a subgroup of U(48): $G_{ext}\subseteq U(48)$. This group
ends up having 7 Higgs fields falling into 4 classes according to the order
of magnitude of the expectation values:

\vspace{0.1cm}

1) The smallest VEV Higgs field plays role of the SM Weinberg--Salam
Higgs field $\phi_{WS}$ having the weak scale value $<\phi_{WS}>=
246~GeV/{\sqrt 2}$.

\vspace{0.1cm}

2)The next smallest VEV Higgs field breaks all families $U(1)_{(B-L)}~$ group,
which is broken at the see--saw scale. This VEV is $<\phi_{(B-L)}>
\sim 10^{12}$ GeV. Such a field is absent in the "old" extended AGUT.

\vspace{0.1cm}

3) The next 4 Higgs fields are W, T, $\xi$ and $\chi$, which have VEVs
of the order of a factor 10 to 50 under the Planck unit. That means
that if intermediate propagators have scales given by the Planck scale,
as it is assumed in AGUT in general, then they will give rise to
suppression factors of the order 1/10 each time they are needed to cause
a transition. The field $\chi$ is absent in the "old" $G_f$--AGUT.
It was introduced in Refs.\ct{37} for the purpose of the study of
neutrinos.

\vspace{0.1cm}

4) The last one, with VEV of the same order as the Planck scale,
is the Higgs field S. It had VEV $<S>=1$ in the "old" extended
AGUT by Froggatt and Nielsen (with $G_f$ group of symmetry), but this VEV
is not equal to unity in the "new" extended AGUT. Therefore there is
a possibility to observe phenomenological consequences of the field S
in the Nielsen--Takanishi model.

Typical fit to the masses and mixing angles for the SM leptons and
quarks in the framework of the $G_{ext}$--AGUT is given in Table II.

In contrast to the "old" extended AGUT by Froggatt--Nielsen (called here
as $G_f$--theory), the new results of $G_{ext}$--theory by Nielsen--Takanishi
 are more encouraging.

We conclude that the $G$--theory, in general,  is successful in describing of
the SM experiment.

\subsection{Multiple Point Principle}

AGUT approach is used in conjuction with the Multiple Point
Principle proposed by D.L.Bennett and H.B.Nielsen \ct{17p}.
According to this principle, Nature seeks a special point -- the Multiple
Critical Point (MCP) -- which is a point on the phase diagram of the
fundamental regulirized gauge theory G (or $G_f$, or $G_{ext}$), where
the vacua of all fields existing in Nature are degenerate having the same
vacuum energy density.
Such a phase diagram has axes given by all coupling constants
considered in theory. Then all (or just many) numbers of phases
meet at the MCP.

MPM assumes the existence of MCP at the Planck scale,
insofar as gravity may be "critical" at the Planck scale.

The philosophy of MPM leads to the necessity
to investigate the phase transition in different gauge theories.
A lattice model of gauge theories is the most convenient formalism also for the
realization of the MPM ideas. As it was mentioned above,
in the simplest case we can imagine our
space--time as a regular hypercubic (3+1)--lattice with the parameter $a$
equal to the fundamental (Planck) scale: $a = \lambda_P = 1/M_{Pl}$.
In general, the lattice results are very encouraging for MPM.

\subsection{AGUT-MPM prediction of the Planck scale values of the
U(1), SU(2) and SU(3) fine structure constants}

The usual definition of the SM coupling constants:
\begin{equation}
  \alpha_1 = \frac{5}{3}\frac{\alpha}{\cos^2\theta_{\ov{MS}}},\quad
  \alpha_2 = \frac{\alpha}{\sin^2\theta_{\ov{MS}}},\quad
  \alpha_3 \equiv \alpha_s = \frac {g^2_s}{4\pi},     \lb{81y}
\end{equation}
where $\alpha$ and $\alpha_s$ are the electromagnetic and SU(3)
fine structure constants, respectively, is given in the Modified
minimal subtraction scheme ($\ov{MS}$).
Here $\theta_{\ov{MS}}$ is the Weinberg weak angle in $\ov{MS}$ scheme.
Using RGE with experimentally
established parameters, it is possible to extrapolate the experimental
values of three inverse running constants $\alpha_i^{-1}(\mu)$
(here $\mu$ is an energy scale and i=1,2,3 correspond to U(1),
SU(2) and SU(3) groups of the SM) from the Electroweak scale to the Planck
scale. The precision of the LEP data allows to make this extrapolation
with small errors (see \ct{33a}). Assuming that these RGEs for
$\alpha_i^{-1}(\mu)$ contain only the contributions of the SM particles
up to $\mu\approx \mu_{Pl}$ and doing the extrapolation with one
Higgs doublet under the assumption of a "desert", the following results
for the inverses $\alpha_{Y,2,3}^{-1}$ (here $\alpha_Y\equiv \frac{3}{5}
\alpha_1$) were obtained in Ref.\ct{17p} (compare with \ct{33a}):
\begin{equation}
   \alpha_Y^{-1}(\mu_{Pl})\approx 55.5; \quad
   \alpha_2^{-1}(\mu_{Pl})\approx 49.5; \quad
   \alpha_3^{-1}(\mu_{Pl})\approx 54.0.
                                                        \lb{82y}
\end{equation}
The extrapolation of $\alpha_{Y,2,3}^{-1}(\mu)$ up to the point
$\mu=\mu_{Pl}$ is shown in Fig.9.

According to AGUT, at some point $\mu=\mu_G < \mu_{Pl}$ (but near
$\mu_{Pl}$) the fundamental group $G$ (or $G_f$, or $G_{\mbox{ext}}$)
undergoes spontaneous breakdown to its diagonal subgroup:
\begin{equation}
      G \longrightarrow G_{diag.subgr.} = \{g,g,g || g\in SMG\},
                                                          \lb{83y}
\end{equation}
which is identified with the usual (low--energy) group SMG.
The point $\mu_G\sim 10^{18}$ GeV also is shown in Fig.9, together with
a region of G--theory, where AGUT works.

The AGUT prediction of the values of $\alpha_i(\mu)$ at $\mu=\mu_{Pl}$
is based on the MPM assumption about the existence of the phase
transition boundary point MCP at the Planck scale, and gives these values
in terms of the corresponding critical couplings $\alpha_{i,crit} $
\ct{2c}-\ct{2k},\ct{17p}:
\begin{equation}
            \alpha_i(\mu_{Pl}) = \frac {\alpha_{i,crit}}{N_{gen}}
                       = \frac{\alpha_{i,crit}}{3}
                \quad{\mbox{for}}\quad i=2,3\mbox{ (also for $i>3$),}       \lb{84y}
\end{equation}
and
\begin{equation}
\alpha_1(\mu_{Pl}) = \frac{\alpha_{1,crit}}{\frac{1}{2}N_{gen}(N_{gen} + 1)}
                   = \frac{\alpha_{1,crit}}{6} \quad{\mbox{for}}\quad U(1).
                                      \lb{85y}
\end{equation}
There exists a simple explanation of the relations (\ref{84y}) and (\ref{85y}).
As it was mentioned above, the group G breaks down at $\mu=\mu_G$.
It should be said that at the very high energies $\mu_G \le \mu \le \mu_{Pl}$
(see Fig.9) each generation has its own gluons, own W's, etc. The breaking
makes only linear combination of a certain color combination of gluons which
exists in the SM below $\mu=\mu_G$ and down to the low energies.
We can say that the phenomenological gluon is a linear
combination (with amplitude $1/\sqrt 3$ for $N_{gen}=3$) for each of the
AGUT gluons of the same color combination. This means that coupling constant
for the phenomenological gluon has a strength that is $\sqrt 3$ times smaller,
if as we effectively assume that three AGUT SU(3) couplings are equal
to each other.
Then we have the following formula connecting the fine structure constants
of G--theory (e.g. AGUT) and low energy surviving diagonal subgroup
$G_{\mbox{diag.subg.}}\subseteq {(SMG)}^3$ given by Eq.(\ref{83y}):
\begin{equation}
\alpha_{\mbox{diag},i}^{-1} = \alpha_{\mbox{1st\, gen.},i}^ {-1} +
\alpha_{\mbox{2nd\, gen.},i}^{-1} + \alpha_{\mbox{3rd\, gen.},i}^{-1}.
                                                      \lb{86y}
\end{equation}
Here i = U(1), SU(2), SU(3), and i=3 means that we talk about the gluon
couplings.
For non--Abelian theories we immediately obtain Eq.(\ref{84y}) from
Eq.(\ref{86y}) at the critical point (MCP).

In contrast to non-Abelian theories, in which the gauge invariance
forbids the mixed (in generations) terms in the Lagrangian of
G--theory, the U(1)--sector of AGUT contains such mixed
terms:
\begin{equation}
\frac{1}{g^2}\sum_{p,q} F_{\mu\nu,\; p}F_{q}^{\mu\nu} =
\frac{1}{g^2_{11}}F_{\mu\nu,\; 1}F_{1}^{\mu\nu} +
\frac{1}{g^2_{12}}F_{\mu\nu,\; 1}F_{2}^{\mu\nu} +
...
+ \frac{1}{g^2_{23}}F_{\mu\nu,\; 2}F_{3}^{\mu\nu} +
\frac{1}{g^2_{33}}F_{\mu\nu,\; 3}F_{3}^{\mu\nu},
\lb{87y}
\end{equation}
where $p,q = 1,2,3$ are the indices of three generations of the AGUT
group $(SMG)^3$. The last equation explains the difference between the
expressions (\ref{84y}) and (\ref{85y}).

It was assumed in Ref.\ct{17p} that the MCP values
$\alpha_{i,crit}$ in Eqs.(\ref{84y}) and (\ref{85y}) coincide with
the triple point values of the effective fine structure
constants given by the generalized lattice SU(3)--, SU(2)-- and U(1)--gauge
theories described by Eqs.(\ref{37}) and (\ref{38}).
Also it was used a natural assumption that the effective $\alpha_{crit}$
does not change its value (at least too much) along the whole bordeline "3"
of Fig.4 for the phase transition "Coulomb--confinement" in the U(1)
lattice gauge theory with the generalized (two parameters) lattice Wilson
action (\ref{38}).

Now let us consider $\alpha_Y^{-1}\,(\approx \alpha^{-1})$ at the point
$\mu=\mu_G\sim 10^{18}$ GeV shown in Fig.9.
If the point $\mu=\mu_G$ is very close to the Planck scale
$\mu=\mu_{Pl}$, then according to Eqs.(\ref{82y}) and (\ref{85y}), we have:
\begin{equation}
         \alpha_{1st\, gen.}^{-1}\approx
    \alpha_{2nd\, gen.}^{-1}\approx \alpha_{3rd\, gen.}^{-1}\approx
    \frac{\alpha_Y^{-1}(\mu_G)}{6}\approx 9,        \lb{88y}
\end{equation}
what is almost equal to the value (\ref{50a}):
$$
            \alpha_{crit.,theor}^{-1}\approx 8
$$
obtained by "Parisi improvement method" (see Fig.5(c)).
This means that in the U(1) sector of AGUT we have $\alpha $ near
the critical point. Therefore, we can expect the existence of MCP
at the Planck scale.

As it was mentioned above, the lattice investigators were not able
to obtain the lattice triple point values of $\alpha_{i,crit}$
(i=1,2,3 correspond to U(1),SU(2) and SU(3) groups) by Monte Carlo
simulation methods.
These values were calculated theoretically by Bennett and Nielsen
in Ref.\ct{17p}. Using the lattice triple point values of
$(\beta_A;\,\beta_f)$ and $(\beta^{lat} ;\,\gamma^{lat})$
(see Fig.3(a,b) and Fig.4), they have obtained $\alpha_{i,crit}$ by the
"Parisi improvement method":
\begin{equation}
    \alpha_{Y,crit}^{-1}\approx 9.2\pm 1,
    \quad \alpha_{2,crit}^{-1}\approx 16.5\pm 1, \quad
    \alpha_{3,crit}^{-1}\approx 18.9\pm 1.                 \lb{89y}
\end{equation}
Assuming the existence of MCP at $\mu=\mu_{Pl}$
and substituting the last results in Eqs.(\ref{84y}) and (\ref{85y}),
we have the following prediction of AGUT \ct{17p}:
\begin{equation}
   \alpha_Y^{-1}(\mu_{Pl})\approx 55\pm 6; \quad
   \alpha_2^{-1}(\mu_{Pl})\approx 49.5\pm 3; \quad
   \alpha_3^{-1}(\mu_{Pl})\approx 57.0\pm 3.
                                                          \lb{90y}
\end{equation}
These results coincide with the results (\ref{82y}) obtained by the
extrapolation of experimental data to the Planck scale
in the framework of the pure SM (without any new particles) \ct{17p},
\ct{33a}.

Using the relation (\ref{25z}), we obtained the result (\ref{26z}),
which in our case gives the following relations:
\begin{equation}
    \alpha_{Y,crit}^{-1} : \alpha_{2,crit}^{-1} : \alpha_{3,crit}^{-1}
           = 1 : \sqrt{3} : 3/\sqrt{2} = 1 : 1.73 : 2.12.
                                                     \lb{91y}
\end{equation}
Let us compare now these relations with the MPM prediction.

For $\alpha_{Y,crit}^{-1}\approx 9.2$  given by the first equation of
(\ref{89y}), we have:
\begin{equation}
 \alpha_{Y,crit}^{-1} : \alpha_{2,crit}^{-1} : \alpha_{3,crit}^{-1}
    = 9.2 : 15.9 : 19.5.                                     \lb{92y}
\end{equation}
In the framework of errors the last result coincides with the
AGUT--MPM prediction (\ref{89y}).
Of course, it is necessary to take into account an approximate description
of confinement dynamics in the SU(N) gauge theories, which was
used in our investigations.

\section{The possibility of the Grand Unification Near the Planck Scale}

We can see new consequences of the extension of $G$--theory, if
$G$--group is broken down to its diagonal subgroup $G_{diag}$, i.e. SM,
not at $\mu_G\sim 10^{18}$ {GeV}, but at $\mu_G\sim 10^{15}$ or $10^{16}$ {GeV}.
In this connection, it is very attractive to consider the gravitational
interaction.

\subsection{"Gravitational finestructure constant" evolution}

The gravitational interaction between two particles
of equal masses M is given by the usual classical Newtonian potential:
\begin{equation}
   V_g = - G \frac{M^2}{r} =
           - \left(\frac{M}{M_{Pl}}\right)^2\frac{1}{r}
                   = - \frac{\alpha_g(M)}{r},              \lb{1x}
\end{equation}
which always can be imagined as a tree--level approximation of quantum
gravity.

Then the quantity:
\begin{equation}
      \alpha_g = \left(\frac{\mu}{\mu_{Pl}}\right)^2     \lb{2x}
\end{equation}
(see also Refs. \ct{10nb,39a,39b}) plays a role of the running "gravitational
finestructure constant" and the evolution of its inverse quantity is presented
in Fig.10 together with the evolutions of $\alpha_{1,2,3}^{-1}(\mu)$ (here we
have returned to the consideration of $\alpha_1$ instead of $\alpha_Y$).

Then we see the intersection of $\alpha_g^{-1}(\mu)$
with $\alpha_1^{-1}(\mu)$ in the region of $G$--theory at the point:
$$
               (x_0, \alpha_0^{-1}),
$$
where
\begin{equation}
      x_0 \approx 18.3,  \quad
       \alpha_0^{-1} \approx 34.4,                   \lb{3x}
\end{equation}
and $\,x = \log_{10}\mu$.

\subsection{The consequences of the breakdown of $G$-theory
at $\mu_G\sim 10^{15}$ or $10^{16}$ GeV}

Let us assume now that the group of symmetry $G$ undergoes the breakdown
to its diagonal subgroup not at $\mu_G\sim 10^{18}$ GeV, but at
$\mu_G\sim 10^{15}$ GeV, i.e. before the intersection of
$\alpha_{2}^{-1}(\mu)$ and $\alpha_{3}^{-1}(\mu)$ at
$\mu\sim 10^{16}$
GeV. Why is it important?

As a consequence of behavior of the function $\alpha^{-1}(\beta)$
near the phase transition point, shown in Fig.5c, we have to expect the
change of the evolution of $\alpha_i^{-1}(\mu)$ in the region $\mu > \mu_G$
shown in Fig.9 by dashed lines. Instead of these dashed lines,
we must see the decreasing of $\alpha_i^{-1}(\mu)$, when they
approach MCP, if this MCP really exists at the Planck scale.

According to Fig.5c, it seems that in the very vicinity of the phase transition point
(i.e. also near the MCP at $\mu=\mu_{Pl}$), we cannot
describe the behavior of $\alpha_i^{-1}(\mu)$ by the one--loop
approximation RGE.

It is well known, that the one--loop approximation RGEs
for $\alpha_i^{-1}(\mu)$ (see for example \ct{1a}) can be
described in our case by the following expression:
\begin{equation}
  \alpha_i^{-1}(\mu) =
  \alpha_i^{-1}(\mu_{Pl}) + \frac{b_i}{4\pi}\log(\frac{\mu^2}{\mu^2_{Pl}}),
                                                \lb{4x}
\end{equation}
where $b_i$ are given by the following values:
$$
   b_i = (b_1, b_2, b_3) =
$$
\begin{equation}
( - \frac{4N_{gen}}{3} -\frac{1}{10}N_S,\,\,
      \frac{22}{3}N_V - \frac{4N_{gen}}{3} -\frac{1}{6}N_S,\,\,
      11 N_V - \frac{4N_{gen}}{3} ).                   \lb{5x}
\end{equation}
The integers $N_{gen},\,N_S,\,N_V\,$ are respectively the numbers
of generations,
Higgs bosons and different vector gauge fields of given "colors".

For the SM we have:
\begin{equation}
       N_{gen} = 3, \quad N_S = N_V =1,                    \lb{6x}
\end{equation}
and the corresponding slopes (\ref{5x}) describe the evolutions of
$\alpha_i^{-1}(\mu)$ up to $\mu = \mu_G$ presented in Fig.10.

But in the region $\mu_G\le \mu \le \mu_{Pl}$, when $G$--theory works,
we have $N_V = 3$ (here we didn't take into account the additional
Higgs fields which can change the number $N_S$), and the one--loop
approximation slopes are almost 3 times larger than the same ones for the SM.
In this case, it is difficult to understand that such evolutions give the
MCP values of $\alpha_i^{-1}(\mu_{Pl})$, which are
shown in Fig.11. These values were obtained by the following way:
\begin{equation}
\begin{array}{l}
\alpha^{-1}_{1,\;\;crit}\equiv  \alpha_1^{-1}(\mu_{Pl}) \approx
6\cdot \frac{3}{5}\alpha_{U(1),crit}^{-1}\approx 13,
\\
\alpha^{-1}_{2,\;\;crit}\equiv\alpha_2^{-1}(\mu_{Pl}) \approx
3\cdot \sqrt{3}\alpha_{U(1),crit}^{-1}\approx 19,
\\
\alpha^{-1}_{3,\;\;crit}\equiv \alpha_3^{-1}(\mu_{Pl}) \approx
3\cdot \frac{3}{\sqrt 2}\alpha_{U(1),crit}^{-1}\approx 24,
\end{array}
\lb{7x}
\end{equation}
where we have used the relation (\ref{25z}) with
\begin{equation}
  \alpha_{U(1),crit} =
  \frac{\alpha_{crit}}{\cos^2\theta_{\ov{MS}}}\approx 0.77\alpha_{crit},
                      \lb{8x}
\end{equation}
taking into account our HMM result (\ref{55by}):
$\alpha_{crit}\approx 0.208$, which coincides with the lattice result
(\ref{47}) and gives:
\begin{equation}
  \alpha_{U(1),crit}^{-1} \approx 3.7.
                                            \lb{9x}
\end{equation}
In the case when $G$--group undergoes the breakdown to the SM not
at $\mu_G\sim 10^{18}$ GeV, but at $\mu_G\sim 10^{15}$ GeV, the
artifact monopoles of non-Abelian  SU(2) and SU(3) sectors of $G$--theory begin
to act more essentially.

According to the group dependence relation (\ref{25z})
(but now we expect that it is very approximate), there
exists, for example, the following estimation
at $\mu_G\sim 10^{15}$ GeV:
\begin{equation}
   \alpha_{U(1)}^{-1}(\mu_G) \sim 7\quad -\quad for\quad
       SU(3)_{1st\,\,gen.}, \,etc.                         \lb{10x}
\end{equation}
which is closer to MCP than the previous value $\alpha_Y^{-1}\sim 9$,
obtained for the AGUT breakdown at $\mu_G\sim 10^{18}$ GeV.

It is natural to assume that $\beta$--functions of SU(2) and SU(3)
sectors of $G$--theory change their one--loop approximation behavior
in the region $\mu > 10^{16}$ GeV and $\alpha_{2,3}^{-1}(\mu)$
begin to decrease, approaching the phase transition (multiple critical)
point at $\mu = \mu_{Pl}$. This means that the assymptotic freedom
of non--Abelian theories becomes weaker near the Planck scale, what can
be explained by the influence of artifact monopoles condensation.
It looks as if these $\beta$--functions have
singularity at the phase transition point and, for example,
can be approximated by the following expression:
\begin{equation}
   \frac{d\alpha^{-1}}{dt} = \frac{\beta(\alpha)}{\alpha}\approx
             A(1 - \frac{\alpha}{\alpha_{crit}})^{-\nu}\quad
              {\mbox{near the phase transition point}}.       \lb{11x}
\end{equation}
This possibility is shown in Fig.11 for $\nu\approx 1$ and $\nu\approx 2.4.$.

Here it is worth-while to comment that such a tendency was revealed
in the vicinity of the confinement region by the forth--loop
approximation of $\beta$--function in QCD (see Ref.\ct{40a}).

\subsection{Does the [SU(5)]$^{\bf 3}$ SUSY unification exist near the Planck scale?}

Approaching the MCP in the region
of $G$--theory ($\mu_G\le \mu \le \mu_{Pl}$), $\alpha_{2,3}^{-1}(\mu)$ show the necessity
of intersection of $\alpha_{2}^{-1}(\mu)$ with $\alpha_{3}^{-1}(\mu)$
at some point of this region if $\mu_G\sim 10^{15}$ or $10^{16}$ GeV (see Fig.11).
If this intersection takes place at the point $(x_0,\,\alpha_0^{-1})$
given by Eq.(\ref{3x}), then we have the unification of all gauge
interactions (including the gravity) at the point:
\begin{equation}
  (x_{GUT};\,\alpha_{GUT}^{-1})\approx (18.3;\,34.4),     \lb{12x}
\end{equation}
where $x = \log_{10}\mu$(GeV).
Here we assume the existence of [SU(5)]$^3$ SUSY unification
with superparticles of masses
\begin{equation}
M\approx 10^{18.3}\, {\mbox{GeV}}.
\lb{13x}
\end{equation}
The scale $\mu_{GUT}=M$, given by Eq.(\ref{13x}), can be considered
as a SUSY breaking scale.

Figures 12(a,b) demonstrate such a possibility of
unification. We have investigated the solutions of joint
intersections of $\alpha_g^{-1}(\mu)$ and all
$\alpha_i^{-1}(\mu)$ at different $x_{GUT}$ with different
$\nu$ in Eq.(\ref{11x}). These solutions exist from $\nu\approx
0.5$ to $\nu\approx 2.5$. The possibility of
poles ($\nu=1,\; 2$) was obtained for QCD beta--function
near the confinement region (see~\cite{40b}).

The unification theory with [SU(5)]$^3$--symmetry was suggested first by S.Rajpoot \ct{40}.

It is essential that in this theory the critical point,
obtained by means of Eqs.(\ref{84y}), (\ref{25z}) and (\ref{9x}),
is given by the following value:
\begin{equation}
\alpha_{5,\; crit}^{-1}\approx 3\cdot
\frac{5}{2}\sqrt{\frac{3}{2}}\;
\alpha_{U(1),\; crit}^{-1} \approx 34.0.
\lb{14x}
\end{equation}
The point (\ref{14x}) is shown in Figures 12(a,b) for
the cases:\\
1. $\nu\approx 1$, $\alpha_{GUT}^{-1}\approx
34.4$, $x_{GUT}\approx 18.3$ and $\mu_G\approx 10^{16}$
shown in Fig.12(a);\\
2. $\nu\approx 2.4$, $\alpha_{GUT}^{-1}\approx
34.4$, $x_{GUT}\approx 18.3$ and $\mu_G\approx 10^{15}$
shown in Fig.12(b).\\
We see that the point (212) is very close to the unification point $\alpha_{GUT}^{-1}\approx 34.4$, given by
Eq.(\ref{12x}). This means that the unified theory, suggested here as the [SU(5)]$^3$ SUSY
unification, approaches the confinement phase at the Planck scale.
But the confinement of all SM particles is impossible in
our world, because then they have to be confined also at
low energies what is not observed in the nature.

It is worth--while to mention that using the Zwanziger formalism for the
Abelian gauge theory with electric and magnetic charges
(see Refs.\ct{41}-\ct{43} and \ct{21p}), the possibility of unification
of all gauge interactions at the Planck scale was considered in Ref.\ct{38}
in the case when unconfined monopoles come to the existence near the Planck
scale. They can appear only in G--theory \ct{44}, because RGEs for monopoles
strongly forbid their deconfinement in the SM up to the Planck scale.
But it is not obvious that they really exist in G--theory. This problem
needs more careful investigations, because our today knowledge about
monopoles is still very poor.

The unified theory, suggested in this Section, essentially differs in its
origin from that one, which was considered in Ref.\ct{38}, because
it does not assume the existence of deconfining monopoles up to the
Planck scale, but assumes the influence of artifact monopoles near the
phase transition (critical) point.

Considering the predictions of this unified theory for the low--energy
physics and cosmology, maybe in future we shall be able to answer the question:
"Does the unification of [SU(5)]$^3$ SUSY type really
exist near the Planck scale?"

In conclusion, it is necessary to comment that in the SR
theory [21--23] all evolutions investigated in this section
have to be considered only as functions of the variable
$\mu=1/r$, where $r$ is a distance. The evolutions are
different if $\mu$ is the energy scale. The explanation is
given in Refs.~[21--23], and we refer to them.

\newpage

\section{Conclusions}

1. In this review we have presented the main ideas of the Nielsen's
Random Dynamics (RD).

2. We have considered the theory of Scale Relativity (SR) by L.Nottale,
which has something in common with RD and predicts the existence
of the fundamental (minimally possible) length in Nature equal to the
Planck length.

3. We have discussed that RD and SR lead to the discreteness of our
space-time, giving rise to the new description of physics at very
small distances.

4. We have reviewed the lattice gauge theories.

5. We have used the approximation of lattice artifact monopoles as fundamental pointlike
particles described by the Higgs scalar fields and considered the dual Abelian Higgs model of scalar monopoles,
or shortly the Higgs Monopole Model (HMM), as a simplest effective dynamics
describing the confinement mechanism in the pure gauge lattice theories.

6. Using the Coleman--Weinberg idea of the RG improvement of the effective
potential, we have considered this potential with $\beta$--functions
calculated in the two--loop approximation.

7. The phase transition between the Coulomb--like and confinement
phases has been investigated in the U(1) gauge theory by the method
developed in the Multiple Point Model (MPM), where degenerate vacua are
considered.

8. We have presented the calculation of the phase transition values of
$\alpha = e^2/4\pi$  and $\tilde \alpha = g^2/4\pi$. Comparing the
result $\alpha_{crit}\approx 0.17$ and ${\tilde \alpha}_{crit}
\approx 1.48$, obtained in the one--loop approximation, with the result
$\alpha_{crit}\approx 0.208$ and ${\tilde \alpha}_{crit}\approx
1.20$,
obtained in the two--loop approximation, we have shown the coincidence
of the critical values of electric and magnetic fine structure constants
calculated in the two--loop approximation of HMM with the lattice result~\cite{10s}:
$\alpha_{crit}\approx 0.20 \pm 0.015$ and ${\tilde \alpha}_{crit}
\approx 1.25 \pm 0.10$.

9. Comparing the one--loop and two--loop contributions to beta--functions,
we have demonstrated the validity of HMM perturbation theory in solution
of the phase transition problem in the U(1) gauge theory.

10. Investigating the phase transition in HMM,
we have pursued two objects: the first aim was to explain the lattice results,
but the second one was to confirm the MPM prediction, according to which
at the Planck scale there exists a Multiple Critical Point (MCP).

11. We have given reviews of the Anti-Grand Unification Theory (AGUT, or
G--theory) and MPM.

12. We have compared the predictions of AGUT and MPM for the Planck scale
value of $\alpha_Y^{-1}(\mu)$ with its lattice and HMM results.
This comparison gives arguments in favour of the existence of MCP
at the Planck scale.

13. We have argued "an approximate universality" of the critical
coupling constants which is very important for AGUT and MPM.

14. We have used the 't Hooft idea~\cite{24p} about the Abelian dominance in the
monopole vacuum of non--Abelian theories: monopoles of the Yang--Mills
theories are the solutions of the U(1)--subgroups, arbitrary embedded
into the SU(N) group.

15. Choosing the Abelian gauge and taking into account that
the direction in the Lie algebra of monopole fields are gauge
independent, we have found an average over these directions
and obtained \un{the group dependence relation} between the phase transition
fine structure constants for the groups $U(1)$ and $SU(N)/Z_N$:
$$
\alpha_{N,crit}^{-1} =
           \frac{N}{2}\sqrt{\frac{N+1}{N-1}} \alpha_{U(1),crit}^{-1}.
$$

16. The significant conclusion for MPM was that in the case of the
AGUT breakdown at $\mu_G\sim 10^{18}$ GeV, using the group
dependence of critical couplings and AGUT prediction of the values of
$\alpha_i(\mu)$ at the Planck scale given in terms of the corresponding
critical couplings $\alpha_{i,crit}$, we have obtained the following
relations:
$$
    \alpha_{Y,crit}^{-1} : \alpha_{2,crit}^{-1} : \alpha_{3,crit}^{-1}
           = 1 : \sqrt{3} : 3/\sqrt{2} = 1 : 1.73 : 2.12.
$$
We have shown that for $\alpha_{Y,crit}^{-1}\approx 9.2$
the last equation gives the following result:
$$
 \alpha_{Y,crit}^{-1} : \alpha_{2,crit}^{-1} : \alpha_{3,crit}^{-1}
    = 9.2 : 15.9 : 19.5,
$$
which confirms the Bennett--Nielsen AGUT--MPM prediction:
$$
    \alpha_{Y,crit}^{-1}\approx 9.2\pm 1,
    \quad \alpha_{2,crit}^{-1}\approx 16.5\pm 1, \quad
    \alpha_{3,crit}^{-1}\approx 18.9\pm 1.
$$

17. We have considered the gravitational interaction between two particles
of equal masses M, given by the Newtonian potential,
and presented the evolution of the quantity:
$$
    \alpha_g = \left(\frac{\mu}{\mu_{Pl}}\right)^2,
$$
which plays a role of the running "gravitational finestructure constant".

18. We have shown that the intersection of $\alpha_g^{-1}(\mu)$
with $\alpha_1^{-1}(\mu)$ occurs in the region of G--theory at the point
$(x_0, \alpha_0^{-1})$ with the following values:
$$
       \alpha_0^{-1} \approx 34.4,\quad \quad
                x_0 \approx 18.3,
$$
where $x = \log_{10}\mu$(GeV).

19. Using the lattice indications of the decreasing of $\alpha_i^{-1}(\mu)$,
when they approach the phase transition points, we have argued that near
the MCP at $\mu=\mu_{Pl}$ the behavior of $\alpha_i^{-1}(\mu)$
cannot be described by the one--loop approximation for RGE.

20. We have calculated the MCP values $\alpha_i^{-1}(\mu_{Pl})$:
$$
 \alpha_{1,\;\;crit}^{-1}\approx 13, \quad
 \alpha_{2,\;\;crit}^{-1}\approx 19,\quad
 \alpha_{3,\;\;crit}^{-1}\approx 24,
$$
using the group dependence relation for critical couplings and our HMM result:
$$
       \alpha_{crit}\approx 0.208.
$$

21. We have considered a new possibility of the breakdown of G--theory
at $\mu_G\sim 10^{15}$ or $10^{16}$ GeV and a possible role of monopoles
at high energies: in this case the abelian artifact monopoles of the non-Abelian
SU(2) and SU(3) sectors of AGUT begin to act more essentially and
the evolutions of $\alpha_{2,3}^{-1}(\mu)$ begin to decrease approaching
the MCP, what leads to the necessity
of intersection of $\alpha_{2}^{-1}(\mu)$ with $\alpha_{3}^{-1}(\mu)$
at some point in the region $\mu_G\le \mu \le \mu_{Pl}$.

22. We have discussed that at present time there exists a possibility
of the intersection between $\alpha_{2}^{-1}(\mu)$ and $\alpha_{3}^{-1}(\mu)$
at the point $x_{GUT}\approx 18.3$ with different $\alpha_{GUT}^{-1}$. In
particular,
$$
   (x_{GUT};\,\alpha_{GUT}^{-1})\approx (18.3;\,34.4)
$$
can be the point of unification of all gauge interactions (including the
gravity).

23. It is natural to assume the existence of the [SU(5)]$^3$ SUSY
unification having superparticles of masses
$$
    M\approx 10^{18.3}\, {\mbox{GeV}},
$$
considered as a SUSY breaking scale.

24. We have demonstrated the possibility of [SU(5)]$^3$ SUSY
unification and the existence of its critical point at
$$
      \alpha_{5,crit}^{-1} = \alpha^{-1}_5(\mu_{Pl})\approx 34.0
$$
what means that {\bf such an unified theory
approaches the confinement phase at the Planck scale, but does not reach it.}

25. We have demonstrated that the Higgs scalar monopoles imitating
the artifact monopoles of our discrete space--time play an essential
role at very high (Planck scale) energies.

\newpage

{\bf ACKNOWLEDGMENTS:}\\

We would like to express special thanks to Holger Bech Nielsen,
whose jubilee is celebrated this year. His contribution to this
review is invaluable.

One of the authors (L.V.Laperashvili) deeply thanks the Niels Bohr Institute
for its wonderful hospitality and financial support, without which
our activity would be impossible.

We are deeply thankful to Prof. K.A.Ter--Martirosyan, whose permanent
attention to our activity greatly stimulated our work.

We are grateful to our colleagues Dr. D.L.Bennett, Dr. C.D.Froggatt,
Dr. N.B.Nevzorov, Y.Takanishi, M.A.Trusov, who always supported us
with their help and fruitful advices.

It is a pleasure to thank all participants of the Bled Workshop
"What comes beyond the Standard Model?" (Bled, Slovenia, 2001)
for the interesting discussions of the problems considered in our review.

One of the author (Larisa Laperashvili) thanks very much Dr. B.G.Sidharth
for his kind invitation of Dr. Larisa to the Fourth International
Symposium (Hyderabad, India), where she has found an excellent physics
by Dr. B.G.Sidharth, Dr. L.Nottale, Dr. A.G.Agnese and other
people, and personally Dr. L.Nottale for his
interesting comments.

Dr. Larisa greatly thanks Dr. L.Nottale and Dr. B.G.Sidharth for
sending her their papers, which were strongly useful for her activity
in connection of this review.
She hopes that in the near future our collaboration will be developed,
together with Prof. Holger Bech Nielsen.

\newpage
\clearpage


\noindent\includegraphics[width=159mm, keepaspectratio=true]{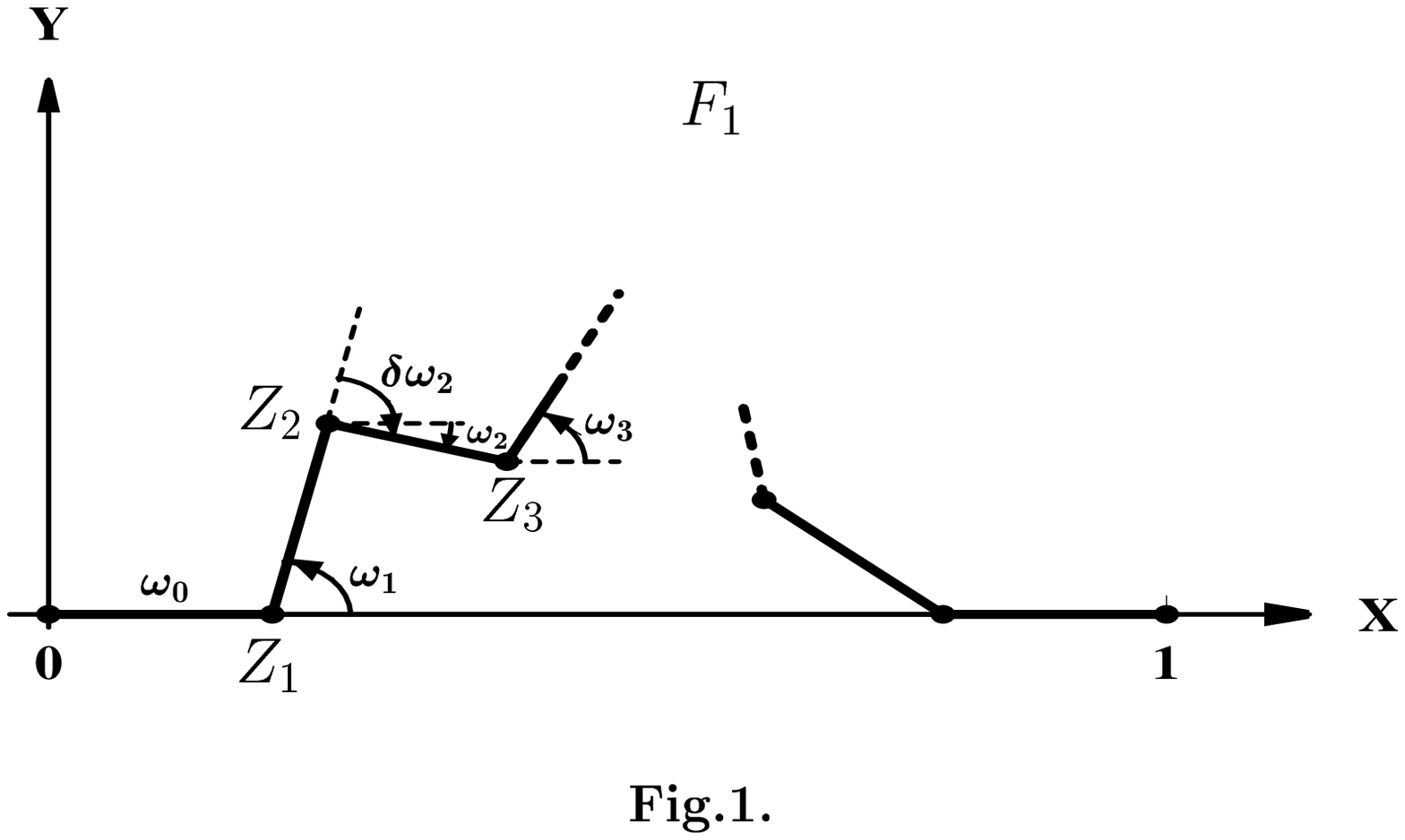}


\vspace*{10mm}

\noindent\includegraphics[width=159mm, keepaspectratio=true]{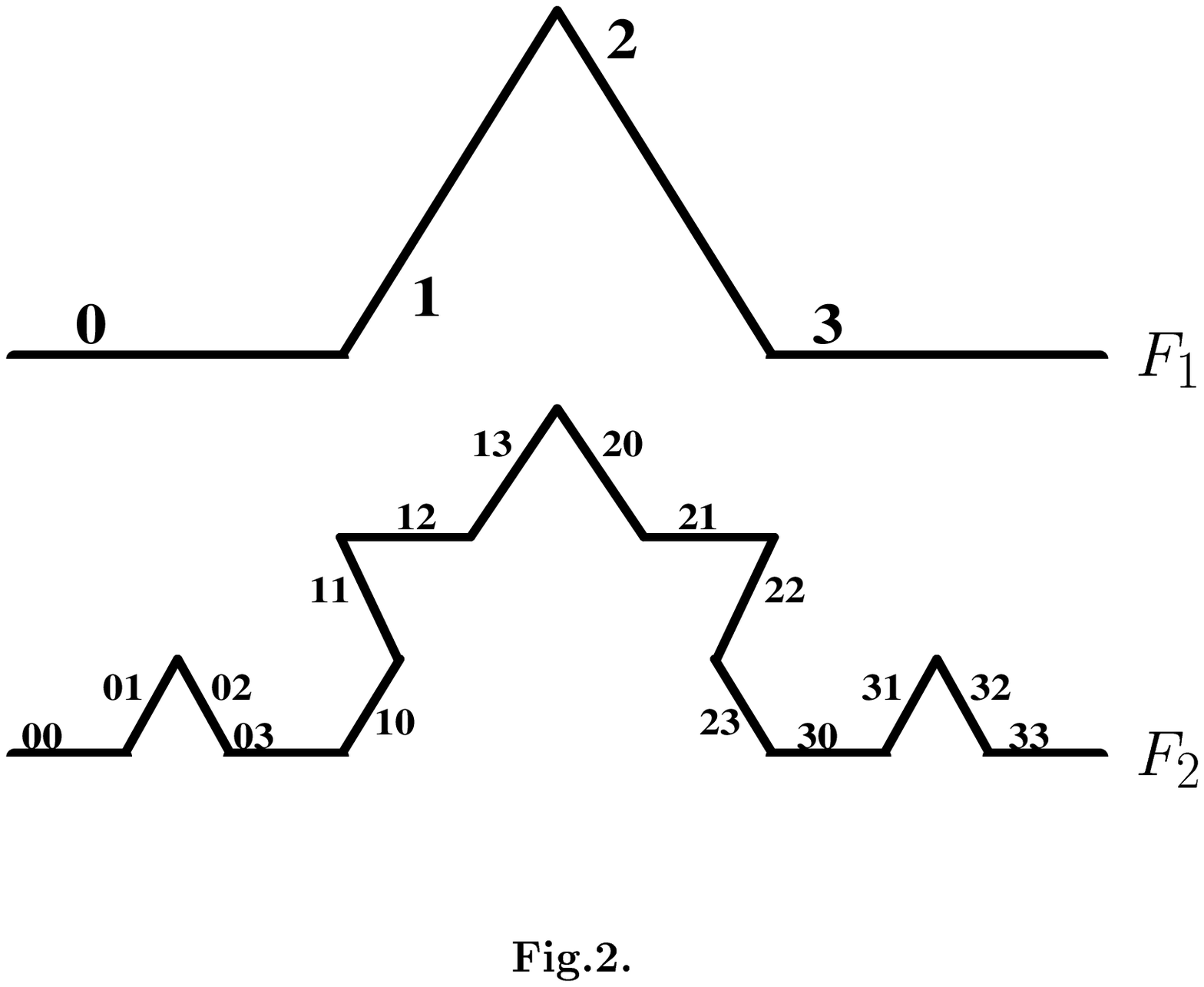}


\noindent\includegraphics[width=159mm, height=235mm]{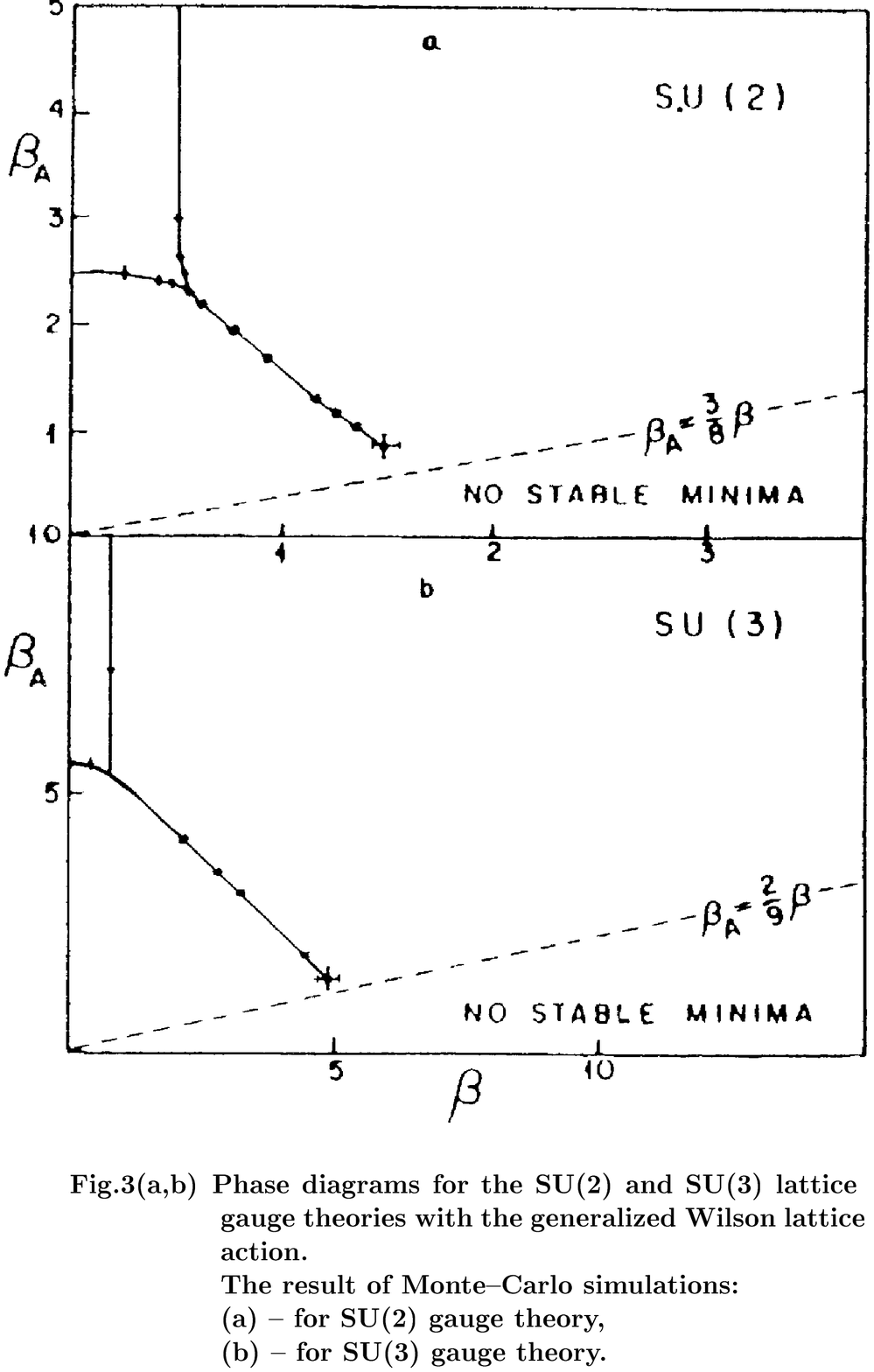}

\newpage
\clearpage

\noindent\includegraphics[width=159mm, keepaspectratio=true]{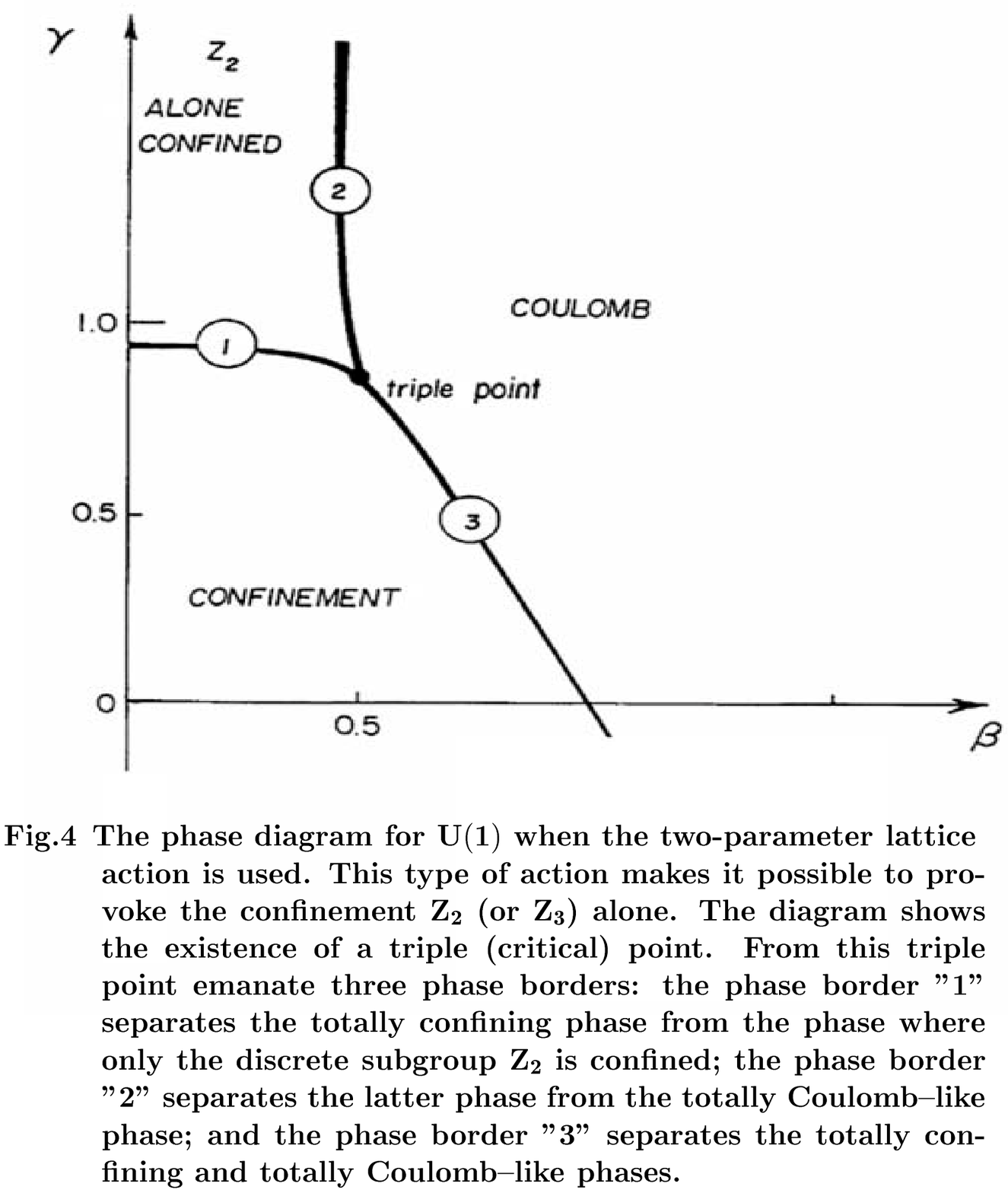}


\noindent\includegraphics[width=159mm, height=235mm]{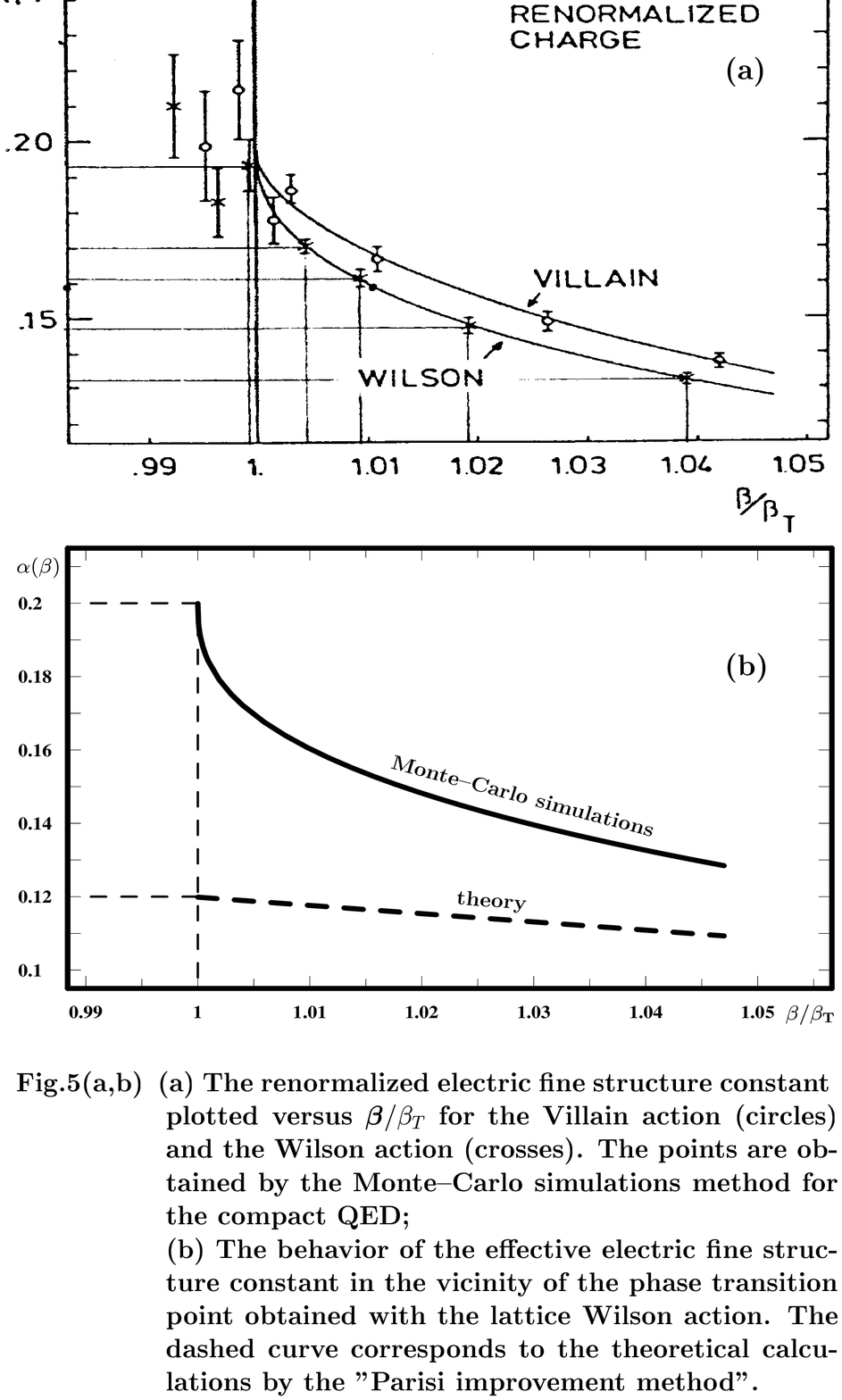}

\newpage
\clearpage

\vspace*{15mm}

\noindent\includegraphics[width=159mm, keepaspectratio=true]{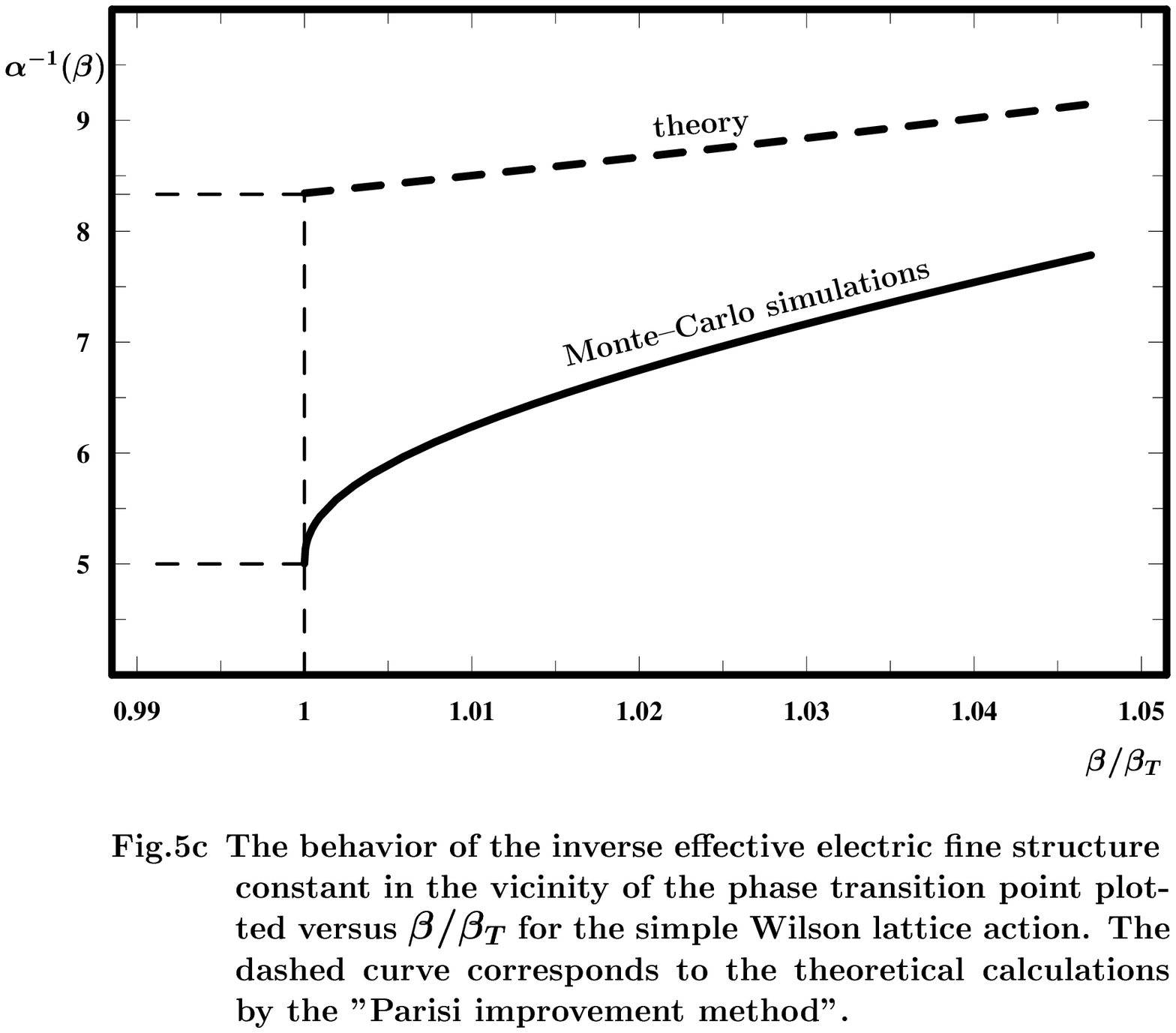}

\newpage
\clearpage

\vspace*{10mm}

\noindent\includegraphics[width=159mm, keepaspectratio=true]{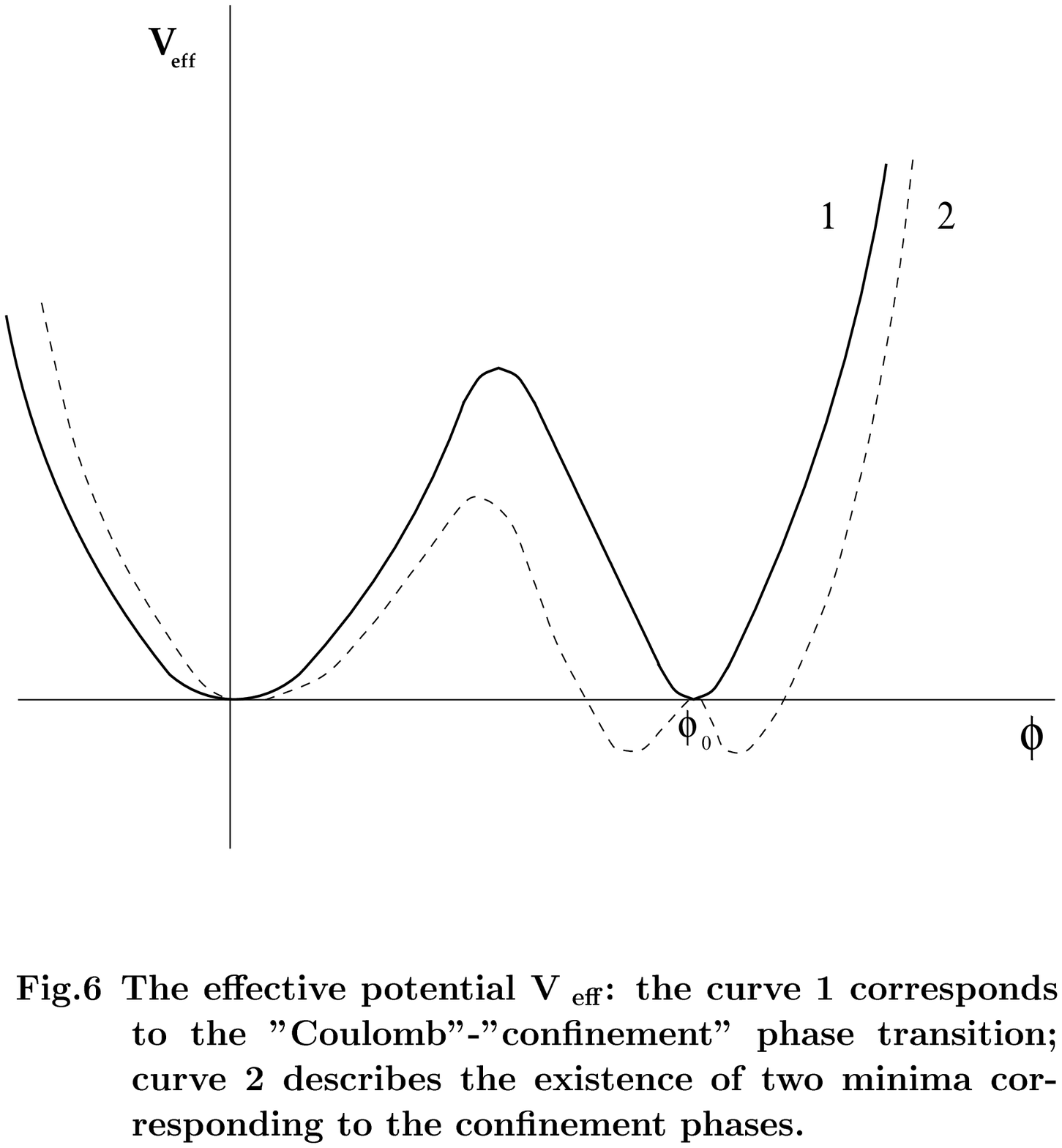}

\newpage
\clearpage

\vspace*{10mm}

\noindent\includegraphics[width=159mm, keepaspectratio=true]{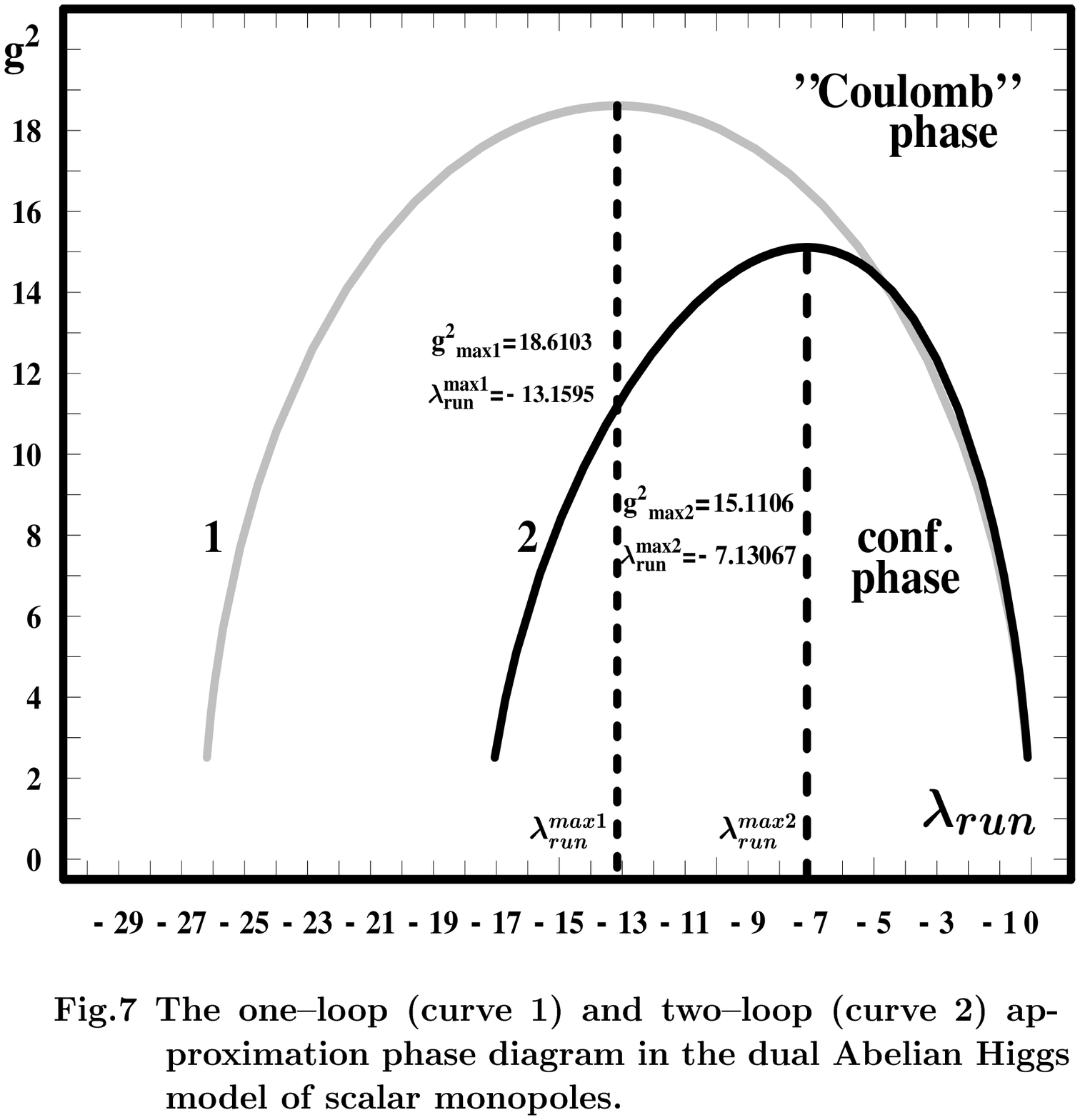}

\newpage
\clearpage

\vspace*{10mm}

\noindent\includegraphics[width=159mm, keepaspectratio=true]{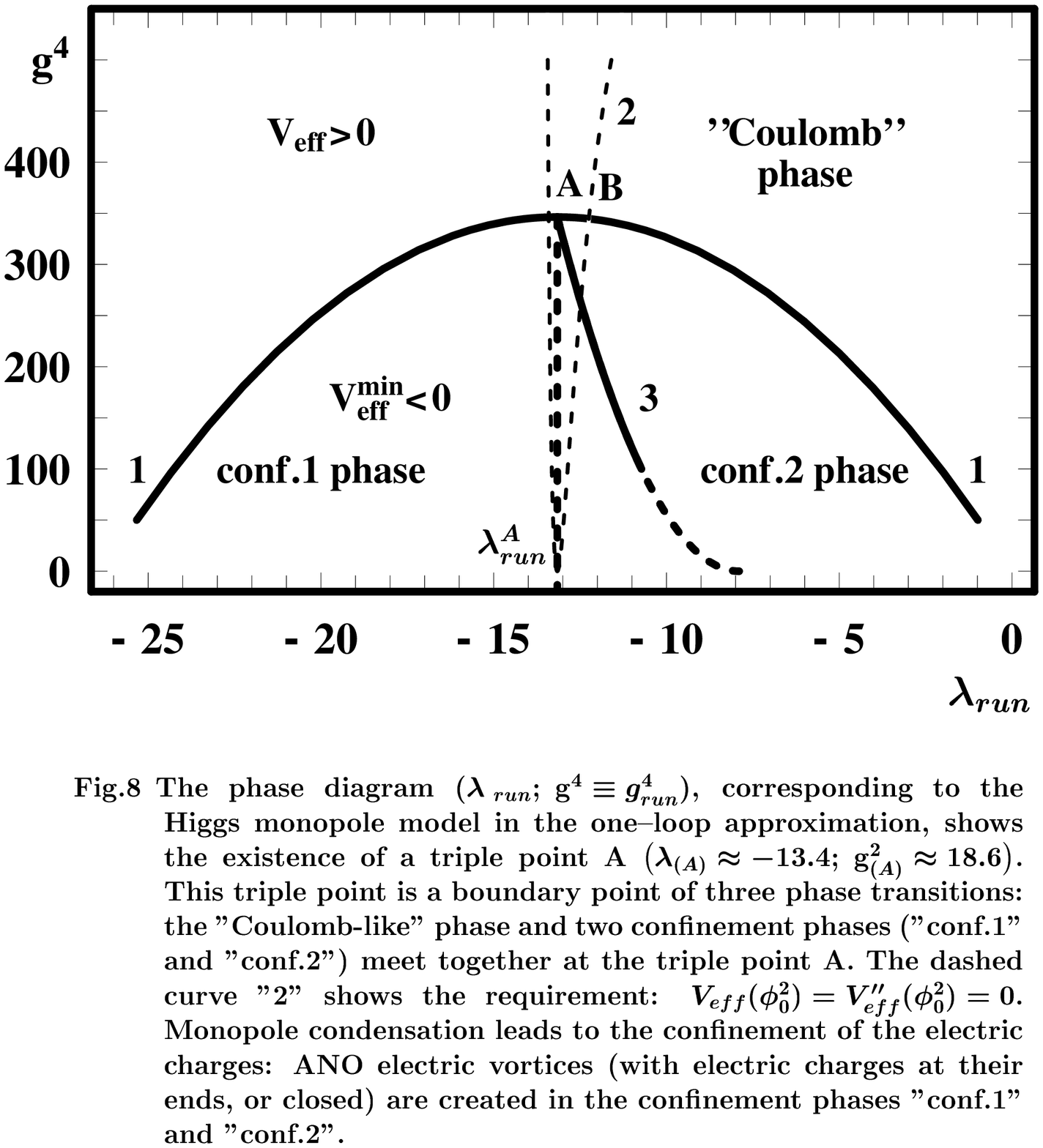}

\newpage
\clearpage

\vspace*{10mm}

\noindent\includegraphics[width=159mm, keepaspectratio=true]{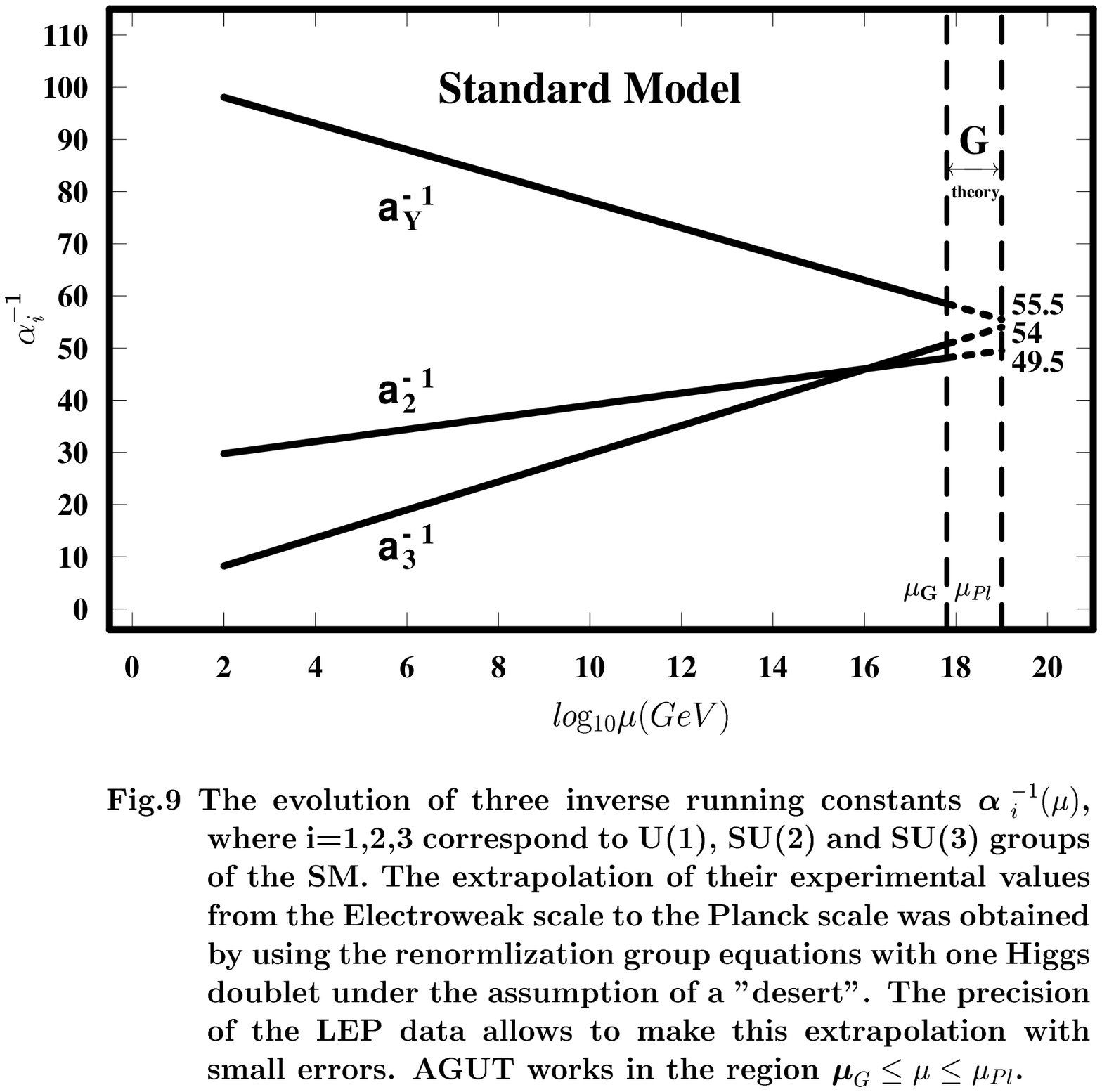}

\newpage
\clearpage

\vspace*{10mm}

\noindent\includegraphics[width=159mm, keepaspectratio=true]{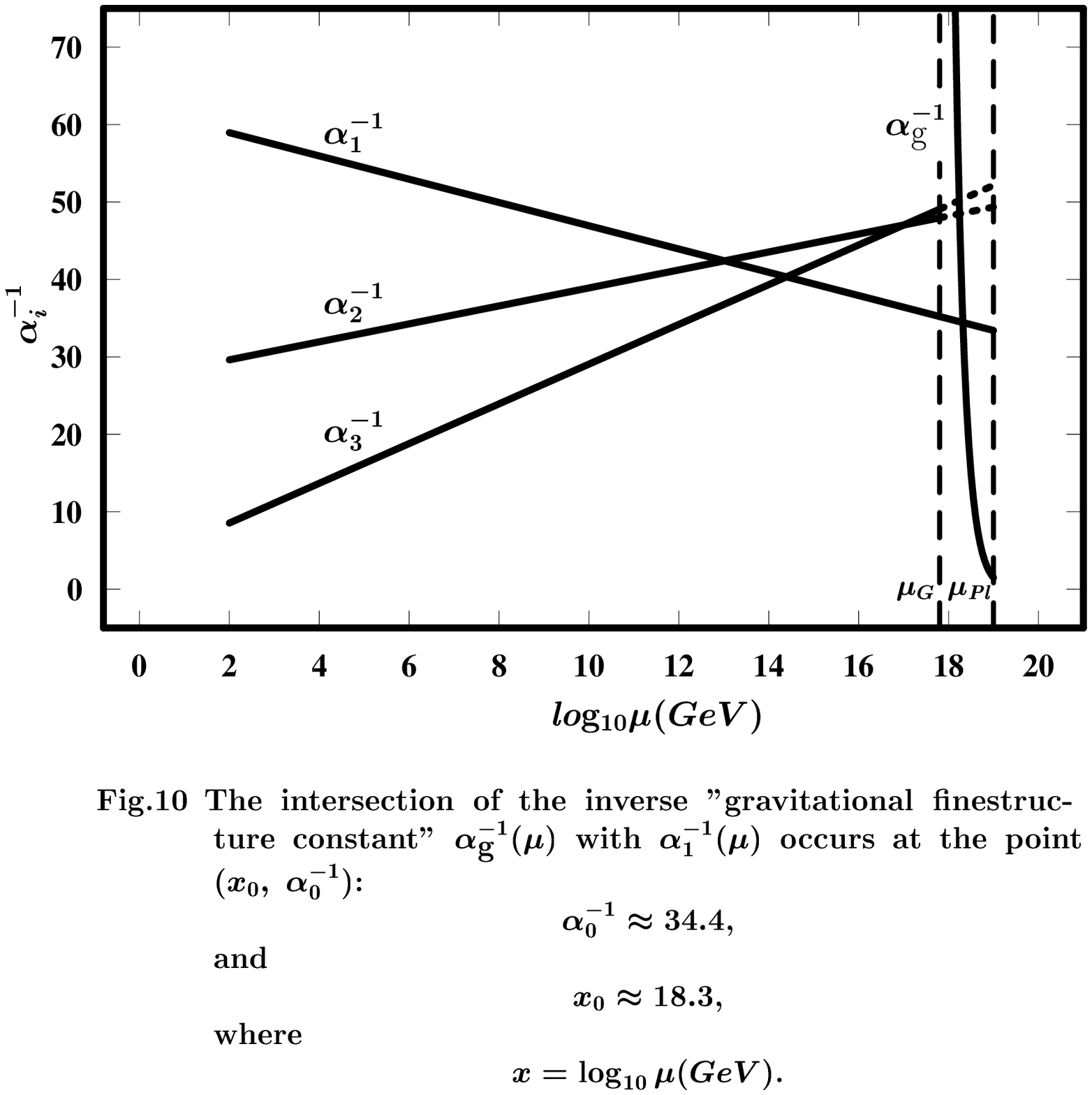}

\newpage


\noindent\includegraphics[width=159mm, height=240mm]{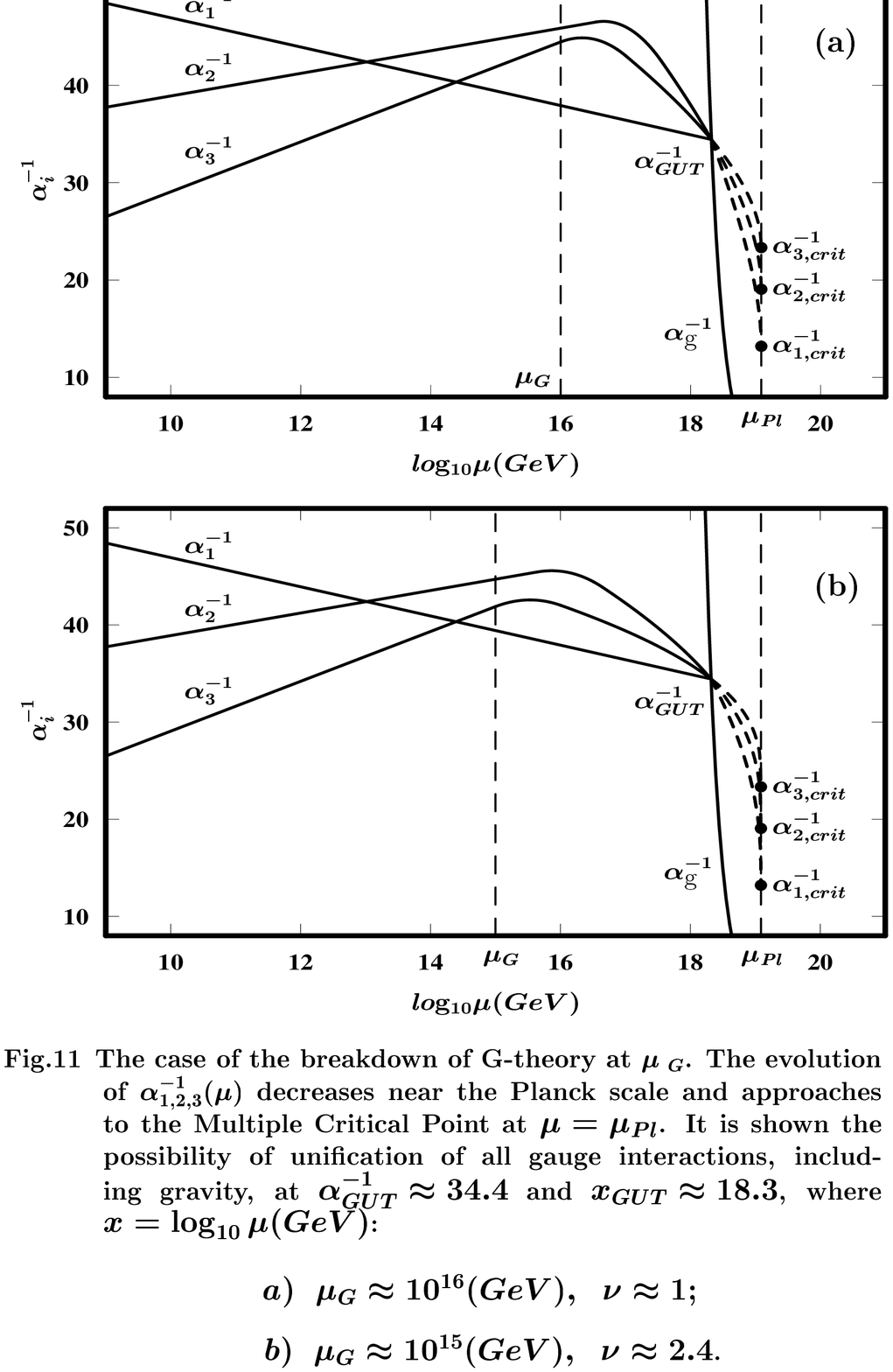}

\newpage


\noindent\includegraphics[width=159mm, height=240mm]{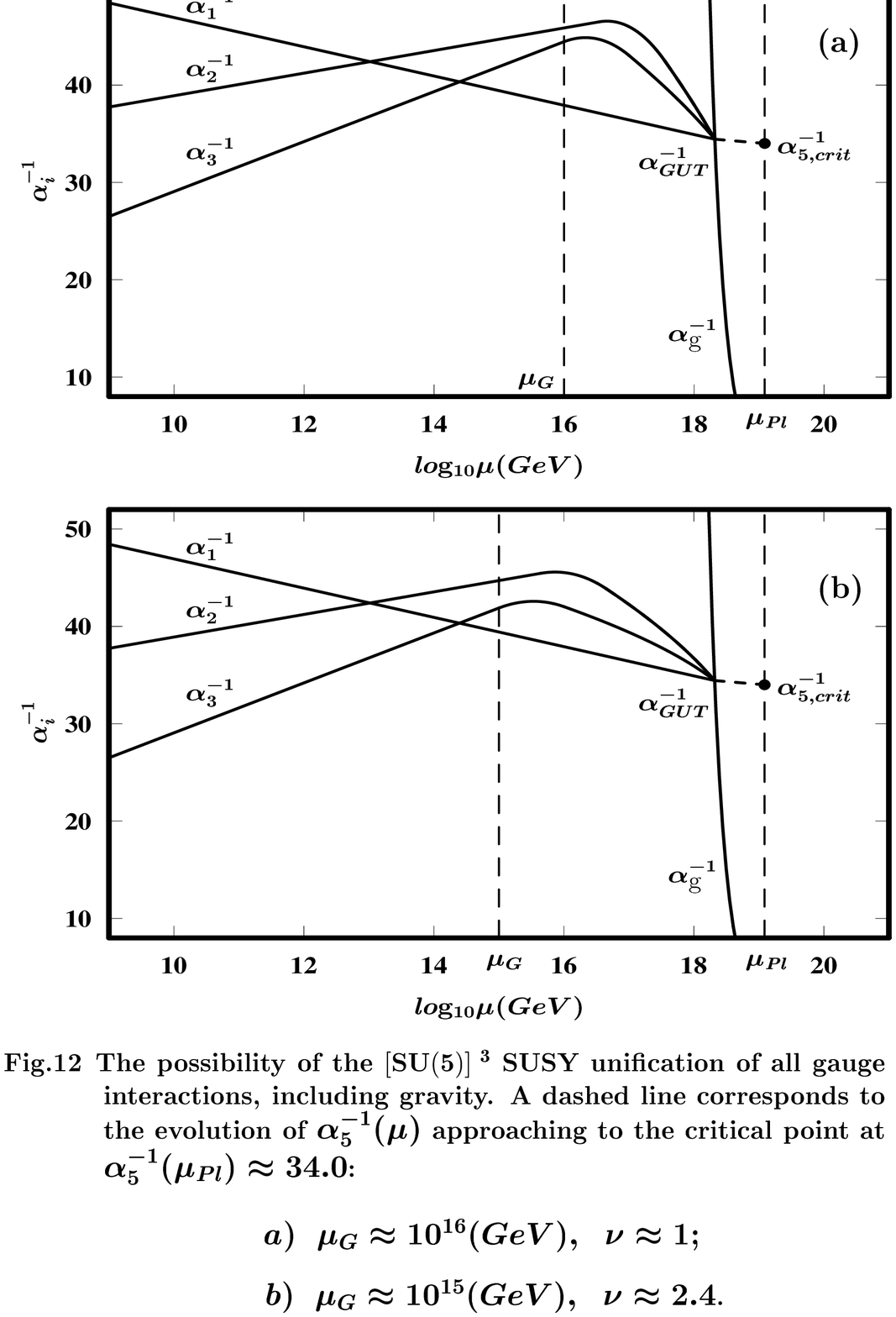}




\newpage

\noindent {\bf Table I.} Best fit to conventional experimental data.
All masses are running masses at 1 GeV except the top quark mass
$M_t$ which is the pole mass.\\

$$
\begin{array}{|c|l|l|}
\hline
&{\mbox {Fitted}}&{\mbox{Experimental}}\\
\hline
m_u & 3.6~{\mbox{MeV}} & 4~{\mbox{MeV}}\\
\hline
m_d&7.0~{\mbox{MeV}}&9~{\mbox{MeV}}\\
\hline
m_e&0.87~{\mbox{MeV}}&0.5~{\mbox{MeV}}\\
\hline
m_c&1.02~{\mbox{GeV}}&1.4~{\mbox{GeV}}\\
\hline
m_s&400~{\mbox{MeV}}&200~{\mbox{MeV}}\\
\hline
m_{\mu}&88~{\mbox{MeV}}&105~{\mbox{MeV}}\\
\hline
M_t&192~{\mbox{GeV}}&180~{\mbox{GeV}}\\
\hline
m_b&8.3~{\mbox{GeV}}&6.3~{\mbox{GeV}}\\
\hline
m_{\tau}&1.27~{\mbox{GeV}}&1.78~{\mbox{GeV}}\\
\hline
V_{us}&0.18&0.22\\
\hline
V_{cb}&0.018&0.041\\
\hline
V_{ub}&0.0039&0.0035\\
\hline
\end{array}
$$

\vspace*{15mm}

\indent {\bf Table II.} Best fit to conventional experimental data in the "new"
AGUT.

$$
\begin{array}{|c|l|l|}
\hline
&{\mbox {Fitted}}&{\mbox{Experimental}}\\
\hline
m_u&3.1~{\mbox{MeV}}&4~{\mbox{MeV}}\\
\hline
m_d&6.6~{\mbox{MeV}}&~9~{\mbox{MeV}}\\
\hline
m_e&0.76~{\mbox{MeV}}&0.5~{\mbox{MeV}}\\
\hline
m_c&1.29~{\mbox{GeV}}&1.4~{\mbox{GeV}}\\
\hline
m_s&390~{\mbox{MeV}}&200~{\mbox{MeV}}\\
\hline
m_{\mu}&85~{\mbox{MeV}}&105~{\mbox{MeV}}\\
\hline
M_t&179~{\mbox{GeV}}&180~{\mbox{GeV}}\\
\hline
m_b&7.8~{\mbox{GeV}}&6.3~{\mbox{GeV}}\\
\hline
m_{\tau}&1.29~{\mbox{GeV}}&1.78~{\mbox{GeV}}\\
\hline
V_{us}&0.21&0.22\\
\hline
V_{cb}&0.023&0.041\\
\hline
V_{ub}&0.0050&0.0035\\
\hline
\end{array}
$$
\vspace{0.2cm}

\noindent All masses are running masses at 1 GeV except the top qurk mass
$M_t$ which is the pole mass.

\noindent The lowest value of ${\tilde \chi}^2$ is $\approx 1.46$.

\newpage

\tableofcontents

\end{document}